\documentclass[useAMS,usenatbib, amsmath, amssymb]{mn2e}
\usepackage{graphicx}
\usepackage{color}

\title[Finite, Intense Accretion Bursts from Tidal Disruption of Stars on Bound Orbits]
{Finite, Intense Accretion Bursts from Tidal Disruption of Stars on Bound Orbits}
\author[K.~Hayasaki, N.~ Stone and A.~Loeb]{Kimitake Hayasaki$^{1,2,3}$\thanks{E-mail:
kimi@kusastro.kyoto-u.ac.jp; kimi@kasi.re.kr}, Nicholas Stone$^{2}$ and Abraham Loeb$^{2}$
\\
$^{1}$Department of Astronomy, Kyoto University, Kitashirakawa-Oiwake-cho, Sakyo-ku, Kyoto 606-8502, Japan\\
$^{2}$Harvard-Smithsonian Center for Astrophysics, 60 GardenStreet, Cambridge, MA02138, USA\\
$^{3}$Korea Astronomy and Space Science Institute, Daedeokdaero 776, Yuseong, Daejeon 305-348, Korea}
\begin{document}

\date{}

\pagerange{\pageref{firstpage}--\pageref{lastpage}} \pubyear{2002}

\maketitle

\label{firstpage}

\begin{abstract}
We study accretion processes for tidally disrupted stars 
approaching supermassive black holes on bound orbits, by performing three
dimensional Smoothed Particle Hydrodynamics simulations with a
pseudo-Newtonian potential. We find that there is a critical value of
the orbital eccentricity below which all the stellar debris remains
bound to the black hole. For high but sub-critical eccentricities,
all the stellar mass is accreted onto the black hole in a finite time,
causing a significant deviation from the canonical $t^{-5/3}$ mass
fallback rate.  When a star is on a moderately eccentric orbit and its
pericenter distance is deeply inside the tidal disruption radius,
there can be several orbit crossings of the debris streams due to
relativistic precession. This dissipates orbital energy in shocks,
allowing for rapid circularization of the debris streams and formation
of an accretion disk. The resultant accretion rate greatly exceeds
the Eddington rate and differs strongly from the canonical rate of $t^{-5/3}$.
By contrast, there is little dissipation due to orbital crossings 
for the equivalent simulation with a purely Newtonian potential. 
This shows that general relativistic precession is crucial for accretion 
disk formation via circularization of stellar debris from stars on moderately eccentric orbits.
\end{abstract}

%
\section{Introduction}
%
Supermassive black holes (SMBHs) larger than $10^5~M_{\odot}$ are now
known to exist ubiquitously in galactic nuclei.  SMBHs in nearby
galaxies can be studied dynamically, whereas in more distant galaxies
only the $\sim 1\%$ of SMBHs undergoing major accretion episodes can
be easily observed.  The tidal disruption and subsequent accretion of
a star by a SMBH, although intrinsically a rare event, is of
observational interest as a way to probe dormant SMBHs in the local
universe, because it can produce a powerful flare in excess of the
Eddington luminosity \citep{cb83, rmj88, ecr89, sq09}.  These events
are also of interest to low-frequency gravitational wave astronomy, as
the stellar dynamical processes which funnel stars into the 
low angular momentum orbits necessary for tidal disruption events
(TDEs) are similar to those which produce the extreme-mass ratio
inspiral of a stellar-mass compact object onto a SMBH \citep{fr76,
wm04, mhl11}.  TDEs have also been considered as part of the means by
which a seed black hole grows into a SMBH \citep{zhr02, mek05,
bcb12}. All of these aspects of stellar tidal disruption have
motivated past studies of TDEs.

The traditional picture of a TDE proceeds as follows: 
a star at large separation ($\sim 1$ pc) approaches a SMBH 
on a nearly parabolic orbit.  After the star is tidally disrupted by the black
hole, half the stellar debris becomes gravitationally bound to the SMBH, because it loses orbital energy inside the tidal radius. The bound
debris falls back, and, after circularizing due to collisional shocks with 
other gas streams, accretes onto the black hole. Kepler's third law implies 
that the mass return rate decays with a -5/3 power of time \citep{rmj88,pse89}, 
asymptotically approaching zero.
This simple analytic picture has been validated to the first order of 
approximation by hydrodynamical simulations \citep{ecr89}, albeit with 
deviations from this law at early times \citep{lg09}. Similar power-law 
behavior for the flare lightcurve is often assumed (i.e. $L \propto \dot{M}$), 
although here the theoretical evidence is less clear \citep{lr11}.

All-sky surveys in the X-ray and UV have so far observed 13 candidate 
tidal disruption flares \citep{ks99,gd99,gj00,ks04,hjp04,dva09,mue10,bjs11,bdn11}.
The observed light curves are in reasonable agreement with the theoretically 
predicted mass fallback rate of $t^{-5/3}$, although some show 
deviations \citep{bdn11} and the number of samples is sufficiently small to 
make detailed testing of theoretical models difficult. 
A smaller number of strong TDE candidates have been found 
at optical wavelengths \citep{vvs11,gs12}. 
Notably, two of the best-sampled TDEs differ strongly from the canonical 
theoretical picture: one possesses relativistic jets \citep{bjs11,zba11}, 
and the other lacks hydrogen lines in its spectra \citep{gs12}.

It has been inferred from observations that the event rate of tidal disruption 
is $\sim 10^{-5} \rm{yr}^{-1}$ per galaxy \citep{djl02}.
This observed rate is in rough agreement with uncertain theoretical rate 
estimates. In the standard scenario, stars are supplied to the SMBH by two body scattering, on a stellar relaxational timescale. 
Stars with angular momentum less than a critical value are in the phase space ``loss cone'' 
and are tidally destroyed on a dynamical time.
Theoretical calculations indicate this rate to be $\sim 10^{-4}-10^{-6} ~{\rm yr}^{-1}$ 
for Milky Way-like galaxies, and that the peak flux into the loss cone comes from 
parsec scales. This motivates the assumption of nearly parabolic orbits 
\citep{mt99, wm04}.

However, there are alternate sources of TDEs distinct from the
standard two-body scattering model. Many of these feed stars to the
SMBH at lower eccentricity.  Our aim here is to quantify through
hydrodynamical numerical simulations how the observable properties of
tidal disruptions of stars on eccentric orbits deviate from the
standard ones.  In \S~\ref{sec:sph-code}, we describe numerical
procedures for simulating TDEs. We then analyze results of our
simulations in \S~\ref{sec:tde}.  In \S~\ref{sec:rates} we consider
nonstandard sources of TDEs and whether they can supply stars on low
eccentricity orbits to a SMBH.  Finally, \S~\ref{sec:sumdis} is
devoted to summary and discussion of our scenario.
 
%
\section{Numerical method}
\label{sec:sph-code}
%
In this section, we describe our procedures for numerically modeling
the tidal disruption of stars on bound orbits. The simulations
presented below were performed with a three-dimensional (3D) SPH code,
which is a particle method that divides the fluid into a set of
discrete \lq\lq fluid elements'' (i.e. particles), and is flexible in
setting various initial configurations. The code is based on a version
originally developed by \cite{benz90a,benz90b,bate95} and has been
extensively used by many papers (e.g., \citet{ato02}; \cite{kh07}). 
The SPH equations with the standard cubic-spline kernel
are integrated using a second-order Runge-Kutta-Fehlberg integrator
with individual time steps for each particle \citep{bate95}, which
results in enormous computational time savings when a large range of
dynamical timescales are involved.  In simulations shown in this
paper, we adopt standard SPH artificial viscosity parameters
$\alpha_\mathrm{SPH}=1$ or $\beta_\mathrm{SPH}=2$, unless otherwise
noted. We have performed two-stage simulations.  We model a star as a
polytropic gas sphere in hydrostatic equilibrium. The tidal disruption
process is modeled by setting the star in motion through the
gravitational field of an SMBH.

\subsection{Formation of polytropes}
\label{sec:polytrope}
%
In our code, the polytrope is composed of an ensemble of gas particles, 
each of which has a mass chosen to be $10^{-5}M_\odot$ with a variable 
smoothing length. The particles are initially uniformly  
distributed in a spherical fashion, with an initial 
temperature $T_{\rm{ini}}=1.2\times10^6{K}$.
The initial spherical gas cloud is allowed to collapse 
under self-gravity using the polytropic equation of state:
\begin{eqnarray}
P=K\rho^{1+1/n},
\label{eq3}
\end{eqnarray}
where $n$ is the polytropic index and $K$ is 
assumed to be kept constant throughout the collapse.
Note that $n=1.5$ corresponds to $\gamma=5/3$. 
The simulations continue over five dynamical times, 
where the dynamical time is defined by
\begin{equation}
\Omega_{*}^{-1}\equiv\sqrt{\frac{r_*^3}{Gm_*}}
\simeq5.1\times10^{-5}
\left(\frac{r_*}{R_\odot}\right)^{3/2}\left(\frac{M_\odot}{m_*}\right)^{1/2}
\,\rm{yr},
\label{eq:tdyn}
\end{equation}
where $G$ is Newton's constant and $r_*$ and $m_*$ are the stellar
radius and mass, respectively.

\begin{figure}
\resizebox{\hsize}{!}{
\includegraphics{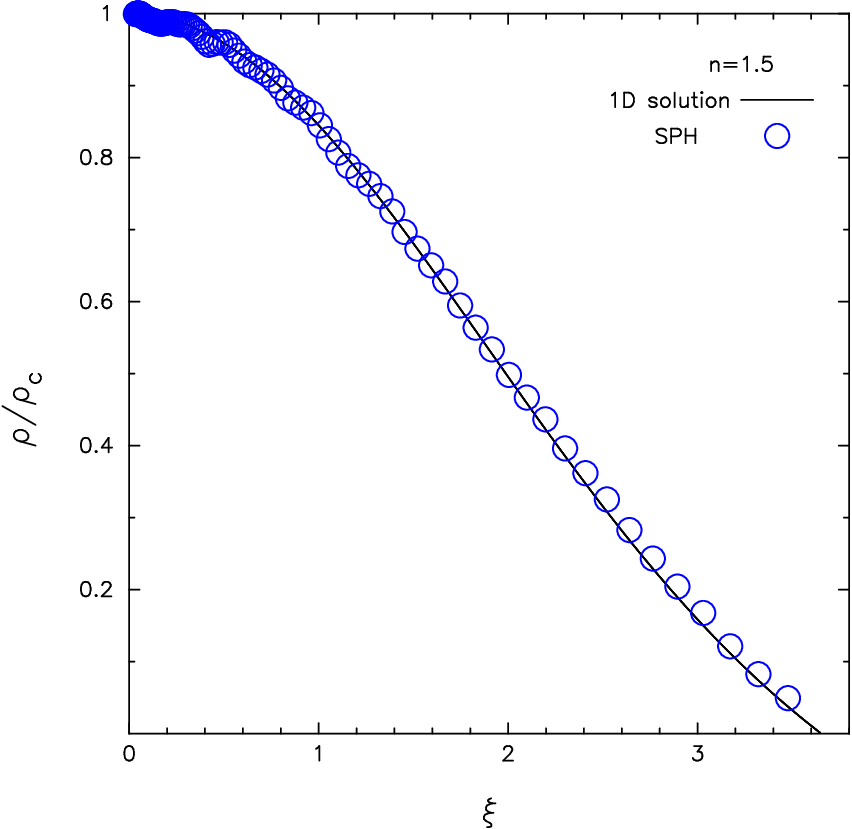}
}
\caption{
The radial density distribution of the polytropic gas sphere with $n=1.5$.
The solid line and circle marks show the solution for the Lane-Emden 
equation with $n=1.5$ and the solution derived from the SPH simulation, 
respectively. The density is normalized by the central density 
$\rho_{\rm{c}}=8.4\,\rm{gcm^{-3}}$ and $\xi=r/r_{\rm{c}}$, 
where $r_{\rm{c}}\simeq0.27R_\odot$.
}
\label{fig:poly}
\end{figure}

Fig.~\ref{fig:poly} shows the radial density profile of the polytropic 
gas sphere at $4.4\Omega^{-1}$, where the magnitude of ratio of 
thermal and gravitational energy becomes $\sim0.5$.
While circle marks show the radial density profile obtained 
from the SPH simulations, the solid line shows the profile 
obtained by numerically solving the Lane-Emden equation:
\begin{eqnarray}
\frac{1}{\xi^2}\frac{d}{d\xi}
\left(
\xi^2\frac{d\theta}{d\xi}
\right)
=-\theta^{n},
\label{eq:lm}
\end{eqnarray}
where we define as dimensionless quantities,
$\xi=r/r_{\rm{c}}$ with
\begin{eqnarray}
r_{\rm{c}}=\left(\frac{K(n+1)}{4\pi{G}}\rho_{\rm{c}}^{1/n-1}\right)^{1/2},
\label{eq:rc}
\end{eqnarray}
and  $\theta^n=\rho/\rho_{\rm{c}}$ with the central density $\rho_{\rm{c}}$.
With boundary conditions that $\theta=1$ and $d\theta/d\xi=0$ at $\xi=0$ 
and $\theta=0$ and $\xi=R_\odot/r_{\rm{c}}$ at the surface of the star for 
1 solar mass sun-type star, we obtained that $\rho_c\simeq8.4\,\rm{g~cm^{-3}}$, 
$r_{\rm{c}}\simeq0.27R_\odot$ and $K\simeq2.5\times10^{15}$. 

%
%

%
\subsection{Initial configuration}
\label{sec:iniconf}
%
A star is tidally disrupted 
when the tidal force of the black hole acting on the star 
is stronger than the self-gravitational force of the star.
The radius where these two forces balance is 
defined as the tidal disruption radius by
\begin{equation}
r_{\rm t}=\left( \frac{M_{\rm BH}}{m_*} \right)^{1/3}r_*,
\label{eq:tdr}
\end{equation}
where $M_{\rm{BH}}$ is the black hole mass.

\begin{figure*}
\resizebox{\hsize}{!}{
\includegraphics{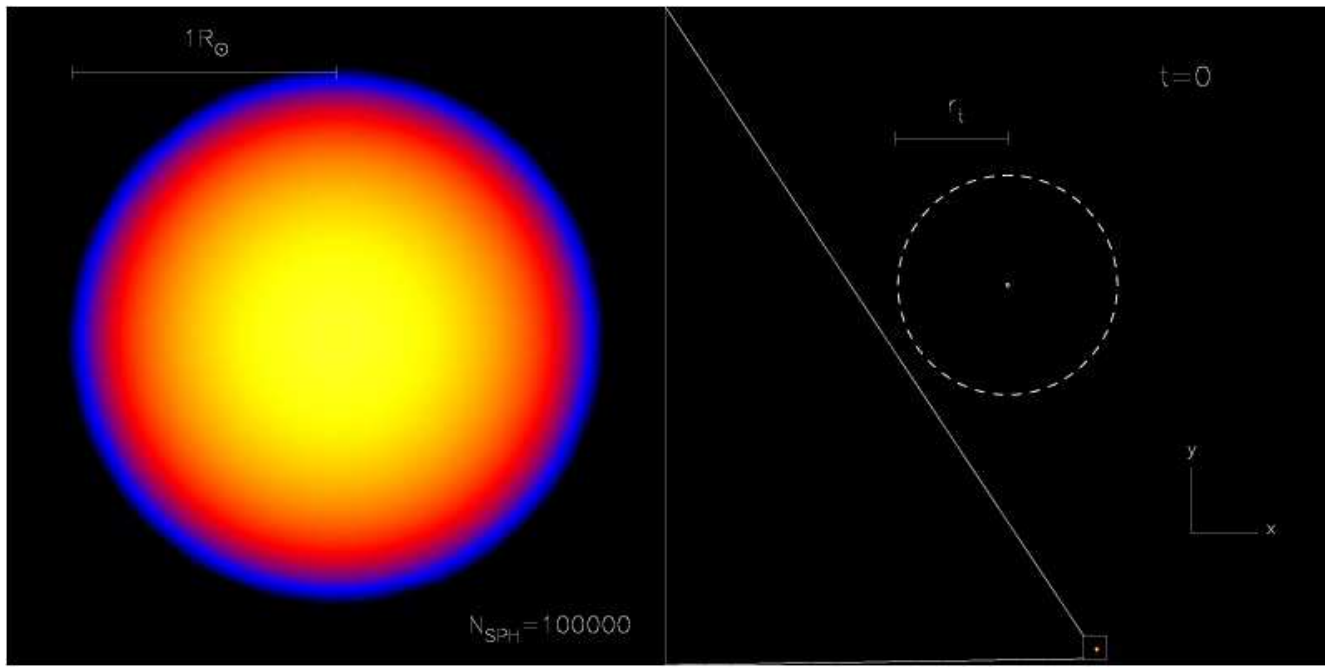}
}
\caption{Initial configuration of our simulations.  The left panel
shows a density map of the polytropic gas sphere with $n=1.5$. Its
radius is $1R_\odot$ and the number of SPH particles $N_{\rm{SPH}}$
are annotated in the right-bottom corner of the panel.  The panel on
the right-hand-side shows the initial position of the star on the
$x$-$y$ plane.
The small square corresponds to the entire left panel, the dashed
circle shows the tidal disruption radius $r_{\rm{t}}$ and the black
hole is positioned at the center.  The simulation run time $t$ in the
right-top corner is normalized by $\Omega_*^{-1}$.  In the right
panel, the star is initially located at
$(r_0\cos\phi_0,r_0\sin\phi_0)$, where $r_0=3r_{\rm{t}}$ and
$\phi_0=-0.4\pi$.  }
\label{fig:iniconf}
\end{figure*}

Fig.~\ref{fig:iniconf} shows a density map of the polytropic gas
sphere at $4.4\Omega_*^{-1}$.  The left panel shows the column density
of polytropic sphere with $n=1.5$ over a range of five orders of
magnitude.  The right panel shows the star-black hole system on the
$x$-$y$ plane, where both axes are normalized by $r_{\rm{t}}$ and the
black hole is put on the origin of the system. The initial position of
the star is given by $\mbox{\boldmath $r$}_0=(r_{\rm 0}\cos\phi_0,
r_{\rm 0}\sin\phi_0, 0)$, where $|\mbox{\boldmath $r$}_0|=r_0$ is the
radial distance from the black hole and $\phi_0$ shows the angle
between $x$-axis and $\mbox{\boldmath $r$}_0$. The star within a small
square is zoomed out to the whole left panel. Note that the stellar
size on the plot has been scaled up by a factor of a few for visual
clarity.

The motion of a test particle in the central SMBH potential $U(r)$
admits two conserved quantities.  The radial velocity and angular
velocity are given by energy conservation and angular momentum
conservation as \citep{lld69}
\begin{eqnarray}
\dot{r}
&=&\sqrt{2(\epsilon-U(r))-\frac{l^2}{r^2}}, \\
\dot{\phi}
&=&\frac{l}{r^2},
\end{eqnarray}
where $\epsilon$ and $l$ are the specific energy and the 
specific angular momentum, respectively.
For bound orbits, 
\begin{eqnarray}
\epsilon&=&
\frac{(r_{\rm{p}}/r_{\rm{a}})^2U(r_{\rm{p}})-U(r_{\rm{a}})}{(r_{\rm{p}}/r_{\rm{a}})^2-1},
\\
l&=&
\sqrt{2r_{\rm{p}}^2(\epsilon-U(r_{\rm{p}}))}=\sqrt{2r_{\rm{a}}^2(\epsilon-U(r_{\rm{a}}))},
\label{eq:enam}
\end{eqnarray}
where $r_{\rm{p}}$ and $r_{\rm{a}}$ are the pericenter distance 
and the apocenter distance, respectively.
Therefore, the initial velocity vector is given by $\mbox{\boldmath $v$}_0= 
(\dot{r}(r_0)\cos\phi_0-r_0\dot{\phi}(\phi_0)\sin\phi_0, 
\dot{r}(r_0)\sin\phi_0+r_0\dot{\phi}(\phi_0)\cos\phi_0, 0)$.

In our simulations, the black hole is represented by a sink particle
with the appropriate gravitational mass $M_{\rm{BH}}$.  All gas
particles that fall within a specified accretion radius are accreted
by the sink particle. We set the accretion radius of the black hole as
equal to the Schwarzshild radius $r_{\rm{S}}=2GM_{\rm{BH}}/c^2$, with
$c$ being the speed of light.

%
%
%
\subsection{Treatment of relativistic precession} 
\label{sec:pseudo}
%
In order to treat approximately the relativistic precession of a test particle 
in the Schwarzschild metric, we incorporate into our SPH code 
the following pseudo-Newtonian potential \citep{wc12}:
\begin{eqnarray}
U(r)
=-\frac{GM_{\rm{BH}}}{r}\left[c_1+\frac{1-c_1}{1-c_2(r_{\rm{S}}/2r)}
+c_3\frac{r_{\rm{S}}}{2r}\right],
\label{eq:wegg}
\end{eqnarray}
where we adopt that $c_1=(-4/3)(2+\sqrt{6})$, $c_2=(4\sqrt{6}-9)$, 
and $c_3=(-4/3)(2\sqrt{6-3})$. Equation~(\ref{eq:wegg}) reduces 
to the Newtonian potential when $c_1=1$ and $c_2=c_3=0$, 
while it reduces to the well-known Paczynski \& Wiita 
pseudo-Newtonian potential \citep{pw80} when $c_1=c_3=0$.  Note that
equation~(\ref{eq:wegg}) includes no higher-order relativistic effects
such as the black hole spin or gravitational wave emission.

\begin{figure}
\resizebox{\hsize}{!}{
\includegraphics{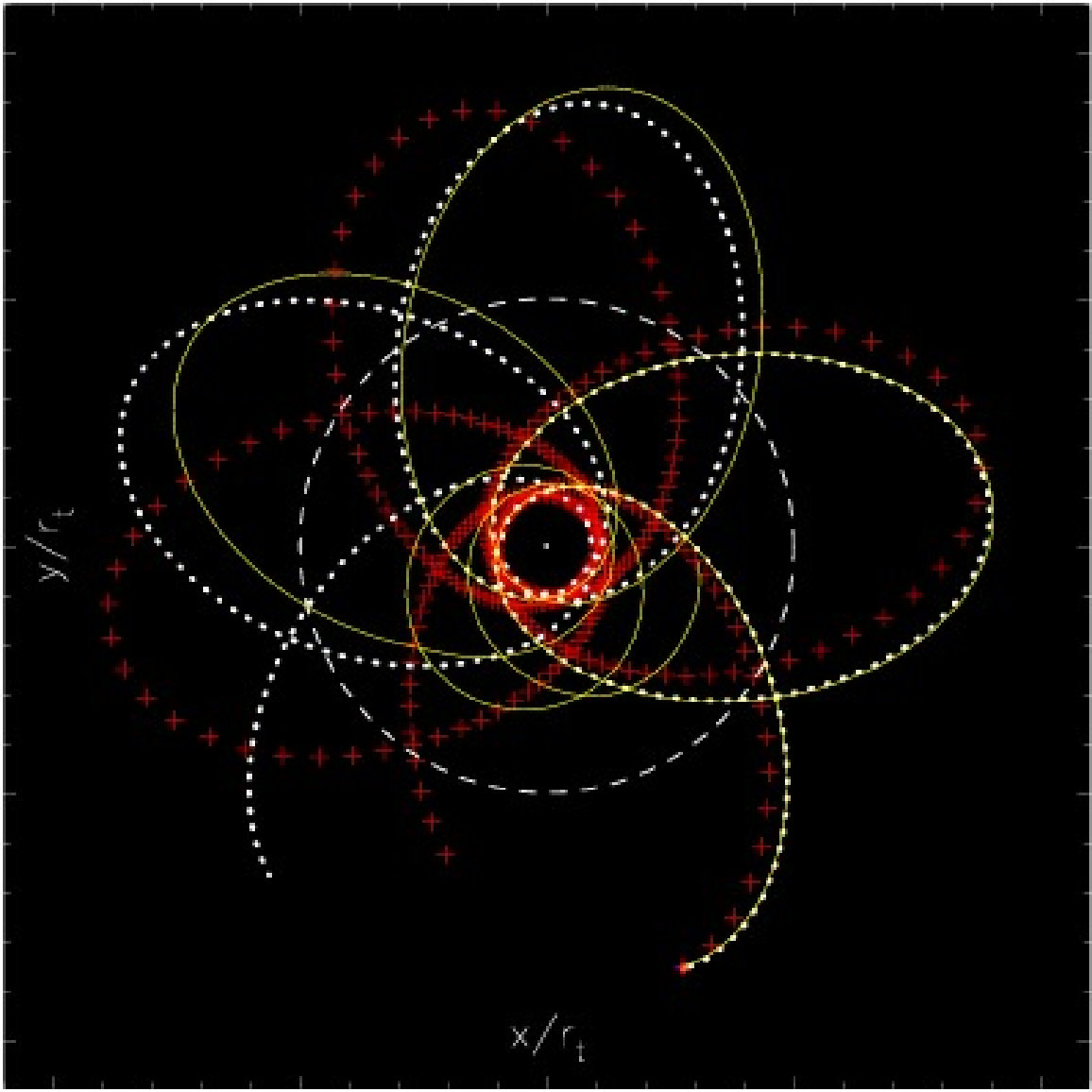}
}
\caption{
Motion of three particles with orbital parameters $(e,\beta)=(0.8,5)$ 
in the $x$-$y$ plane. Each axis is normalized by $r_{\rm{t}}$ given 
by equation~(\ref{eq:tdr}). The central point and dashed circle 
represent the black hole and tidal disruption radius, respectively. 
The small crosses, dotted line, and solid line show the geodesics 
in the Schwarzshild metric, orbits of a test particle under the 
pseudo-Newtonian potential given by equation~(\ref{eq:wegg}), 
and orbits of a SPH particle in the simulation of Model~2a 
(see text of \S~\ref{sec:diskform}), respectively. 
}
\label{fig:grweggsim}
\end{figure}

Fig.~\ref{fig:grweggsim} tests how accurately the SPH particles
composing the star follow test particle motion in the Schwarzschild
metric.  Specifically, the figure shows the motion of three particles
orbiting the black hole with $(e,\beta)=(0.8,5)$, where $\beta=r_{\rm
t}/r_{\rm p}$ is the penetration factor which determines how deeply
the star plunges into the black hole potential inside the tidal
disruption radius.  The central point and dashed circle show the black
hole and the tidal disruption radius, respectively.  The small
crosses, dotted line, and solid line represent the geodesics of the
Schwarzschild metric, orbits of the test particle under the
pseudo-Newtonian potential, and those of a SPH particle in the
simulation of Model~2a, respectively.

\begin{figure}
\resizebox{\hsize}{!}{
\includegraphics{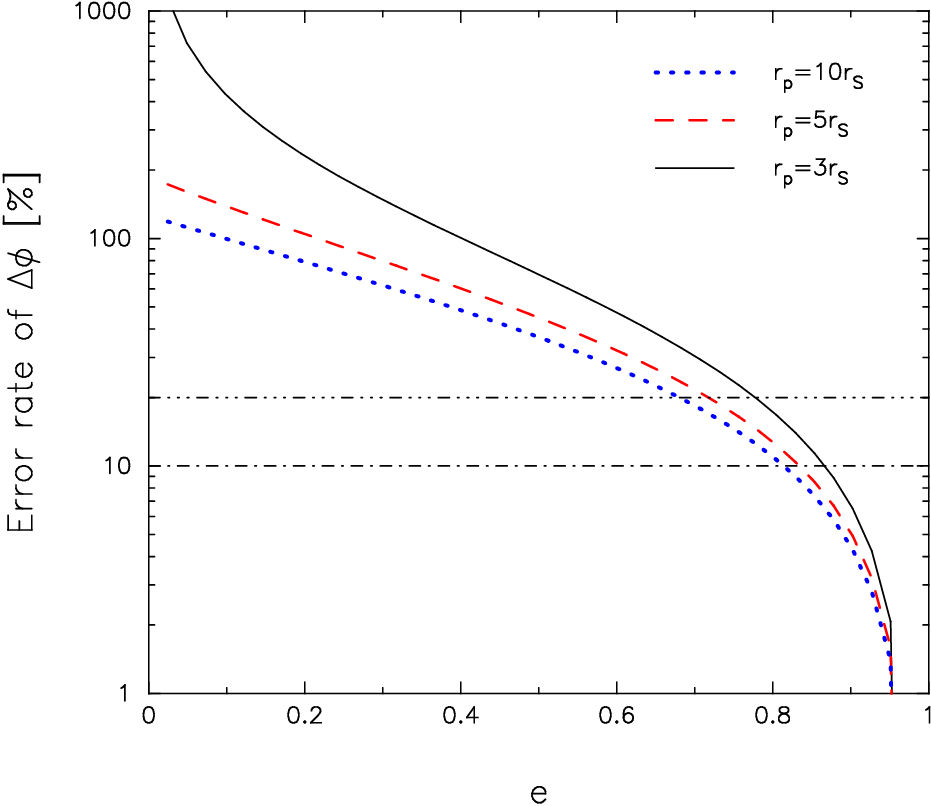}
}
\caption{
Dependence of error rate of precession angle $\Delta\phi$
on the eccentricity of a test particle moving under the pseudo-Newtonian
potential given by equation~(\ref{eq:wegg}).
The solid line, dashed line, and dotted line show 
error rates of an orbit with $r_{\rm{p}}=3r_{\rm{S}}$, $r_{\rm{p}}=5r_{\rm{S}}$,
and $r_{\rm{p}}=10r_{\rm{S}}$, respectively.
The horizontal dash-dotted line and dash-three-dotted line 
show the error cutoffs of $10\%$ and $20\%$, respectively.
}
\label{fig:error_dphi}
\end{figure}

During the first one and a half orbits, the test particle is in
good agreement with the corresponding geodesic.  The subsequent
deviation from the geodesic is due to the relatively low eccentricity
and high $\beta$, since equation~(\ref{eq:wegg}) is tailored to
parabolic orbits with lower $\beta$ \citep{wc12} (see also Fig.~\ref{fig:error_dphi}). 
The first three and half orbits of the test particle are 
in rough agreement with corresponding orbits of the
SPH particle.  Subsequent large deviations originate from orbital
circularization via shock-induced energy dissipation (see
Section~\ref{sec:diskform} in detail), and although our model becomes
unreliable at this point, it is clear that stream crossing is leading
to rapid energy dissipation.

Fig.~\ref{fig:error_dphi} shows the dependence of the error rate 
on the eccentricity of a test particle moving under the pseudo-Newtonian 
potential given by equation~(\ref{eq:wegg}). 
The error rate is defined by $(\Delta\phi_{\rm{GR}}-\Delta\phi_{\rm{PN}})/\Delta\phi_{\rm{GR}}\times100$,
where $\Delta\phi_{\rm{GR}}$ and $\Delta\phi_{\rm{PN}}$ are the precession angles 
corresponding to equation (2) and (4) of \citet{wc12}, respectively, 
with a given specific angular momentum, specific orbital energy, 
and equation~(\ref{eq:wegg}).
The solid line, dashed line, and dotted line show 
error rates of an orbit with $r_{\rm{p}}=3r_{\rm{S}}\,(\beta\approx7.9)$, 
$r_{\rm{p}}=5r_{\rm{S}}\,(\beta\approx4.7)$,
and $r_{\rm{p}}=10r_{\rm{S}}\,(\beta\approx2.4)$, respectively,
where $r_{\rm{t}}\approx23.6r_{\rm{S}}$ is adopted.
The error rate increases as the eccentricity is lower and the penetration factor 
is higher. From the figure, the error rate is estimated to be roughly $15\%$ for $e=0.8$ with $\beta=5$,
while it is less than $1\%$ for $e\ga0.95$ with arbitrary $\beta$.

%
%
%
%
%
%

%
\subsection{Numerical models}
\label{sec:models}
%
We have performed eight simulations of tidal disruption events 
with different parameters.
The common parameters through all of simulations are following: 
$m_*=1M_\odot$, $r_*=1R_\odot$, $M_{\rm{BH}}=10^6M_\odot$, 
$\phi_0=-0.4\pi$, and $\gamma=5/3$. 
The total number of SPH particles used in each simulation are 
$10^5$, and the termination time of each simulation is $4\Omega_{*}^{-1}$.
Table 1 summarizes each model. 
Model 1a shows the standard TDE under the Newtonian potential, 
while Model~1b has the same simulation parameters as Model~1a, except that
the star moves under the pseudo-Newtonian potential given by equation~(\ref{eq:wegg}).
Model 1c and 1d have the same parameters as Model 1a and Model1b, respectively, 
but for $\beta=5$.
Model~2a has the same parameters as Model~1d except 
that the star is on an eccentric orbit, with $e=0.8$. 
Model~2b is the same parameters as Model~2a except that 
the star moves under the Newtonian potential.
Model~3a has the same parameters as Model~2a but for $e=0.98$ and $\beta=1$, and
Model~3b has the same parameters as Model~3a but for $\beta=5$.
%
%
\begin{table*}
 \centering
  \caption{
The first column shows each simulated scenario. 
The second, third, fourth, and fifth columns are 
the penetration factor $\beta=r_{\rm{p}}/r_{\rm{T}}$, the initial orbital 
eccentricity $e_*$, the initial semi-major axis $a_*$, and the radial 
distance between the black hole and the initial position of the star, 
respectively. 
The last column describes the remark for each model.
}
  \begin{tabular}{@{}ccccccccccl@{}}
  \hline
Model 
& {$\beta=r_{\rm{t}}/r_{\rm{p}}$} 
& {$e_*$} & {$a_*\,[r_{\rm{t}}]$} & {$r_0\,[r_{\rm t}]$} & {Remarks} \\
 \hline
1a &  $1$ & $1.0$  & $-$ & $3$  & Newtonian \\
1b &  $1$ & $1.0$  & $-$ & $3$  & Pseudo-Newtonian \\
1c &  $5$ & $1.0$  & $-$ & $3$  & Newtonian \\
1d &  $5$ & $1.0$  & $-$ & $3$  & Pseudo-Newtonian \\
2a &  $5$ & $0.8$  & $1.0$ & $1.8$ & Pseudo-Newtonian \\
2b &  $5$ & $0.8$  & $1.0$ & $1.8$ & Newtonian \\
3a &  $1$ & $0.98$  & $50.0$ & $3$ & Pseudo-Newtonian \\
3b &  $5$ & $0.98$  & $10.0$ & $3$ & Pseudo-Newtonian \\
\hline
\end{tabular}
\end{table*}

%
\section{Tidal disruption of stars on bound orbits}
\label{sec:tde}
%
We first describe the evolution of a tidally disrupted star for
standard TDEs ($e=1$).  Next, the process of tidal disruption for a
star on a fairly eccentric orbit ($e=0.98$) is presented.  Finally,
accretion disk formation in relatively low eccentricity TDEs ($e=0.8$) is
presented.

As an approaching star enters into the tidal disruption radius, 
its fluid elements become dominated by the tidal force of the black hole, 
while their own self-gravity and pressure forces become relatively negligible. 
The tidal force then produces a spread in specific energy of the stellar debris
 \begin{equation}
 \Delta\epsilon\approx \frac{GM_{\rm{BH}}r_*}{r_{\rm t}^2}.
 \label{eq:spreade}
 \end{equation}
The total mass of the stellar debris is defined with the differential mass distribution 
$m(\epsilon)\equiv dM(\epsilon)/d\epsilon$ by
\begin{equation}
M(\epsilon)\equiv\int_{-\infty}^{\infty}m(\epsilon^{'})d\epsilon^{'}.
\label{eq:totm}
\end {equation}
When a star is disrupted from a parabolic orbit, 
$m(\epsilon)$ will be centered on zero and distributed 
over $-\Delta \epsilon\le\epsilon\le\Delta\epsilon$.

Since the stellar debris with negative specific energy 
is bounded to the black hole, it returns to pericenter and 
will eventually accrete onto the black hole. 
If its specific energy is approximately equal
to the specific binding energy, $\epsilon\approx-GM_{\rm{BH}}/2a$ 
with the semi-major axis of the stellar debris $a$,
the mass fallback rate is then defined by (e.g., \cite{ecr89})
\begin{equation}
\frac{dM}{dt}=\frac{dM(\epsilon)}{d\epsilon}
\left|\frac{d\epsilon}{dt}\right| \,\,(\epsilon<0),
\end{equation}
where
\begin{equation} 
\frac{d\epsilon}{dt}=-\frac{1}{3}(2\pi{GM_{\rm{BH}}})^{2/3}t^{-5/3}.
\label{eq:et}
\end{equation}
This is derived from the relationship between the orbital period 
and the specific energy:
\begin{equation}
t = \frac{\pi}{\sqrt{2}}\frac{GM_{\rm{BH}}}{\epsilon^{3/2}},
\label{eq:te}
\end{equation}
from Kepler's third law. 
The standard $dM/dt\propto t^{-5/3}$ scaling then arises 
with the assumption that $dM/d\epsilon$ is constant, 
or at least asymptotes to a constant value at late times.
From equation~(\ref{eq:te}), the orbital period of the most tightly 
bound orbit ($t_{\rm{min}}$) and that of the most loosely bound orbit 
($t_{\rm{max}}$) are given by
\begin{equation}
t \rightarrow
  \left\{ \begin{array}{ll}
    t_{\rm{min}}=
    (\pi/\sqrt{2})\left(M_{\rm{BH}}/m_*\right)^{1/2}\Omega_{*}^{-1} 
    & (\epsilon=\Delta\epsilon) \\
    t_{\rm{max}}=
    \infty & (\epsilon=0)
  \end{array} \right..
\end{equation}

\begin{figure*}
\resizebox{\hsize}{!}{
\includegraphics*[width=3mm]{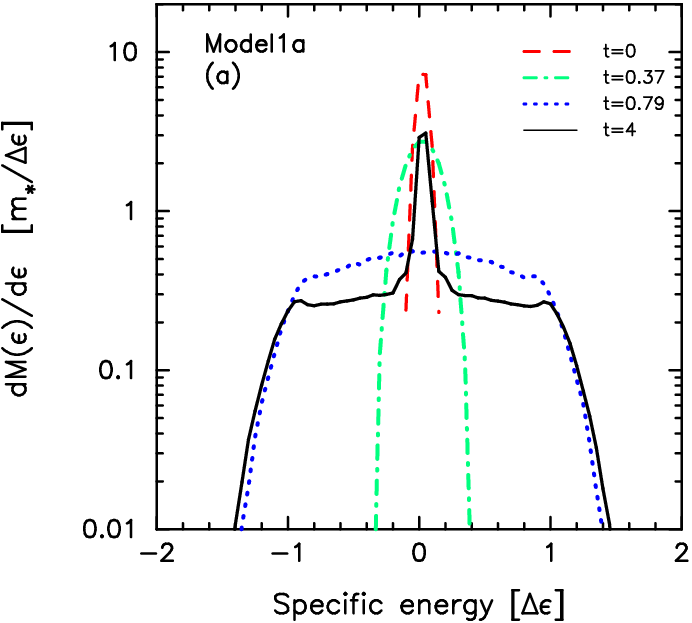}
\includegraphics*[width=3mm]{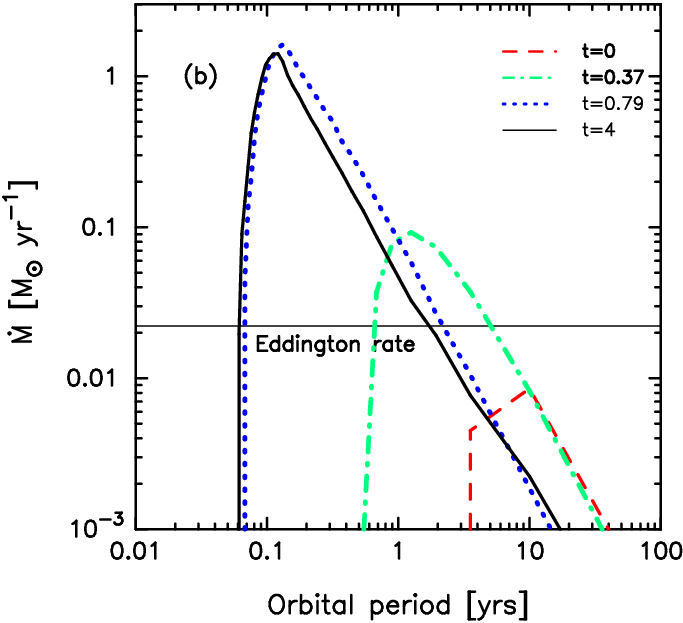}
}
\caption{
Evolution of differential mass distributions over specific energy of stellar debris and 
their corresponding mass-fall back rates, in Model~1a. In panel~(a), the dashed line (red), 
dash-dotted (fairly green), dotted line (blue) and solid line (black) show the differential 
mass distributions measured at $t=0$, $t=0.37$, $t=0.79$, and $t=4$, respectively. 
The energy is measured in units of $\Delta\epsilon$ given by equation~(\ref{eq:spreade}).
In panel~(b), the dash line (red), dashed-dotted line (fairly green), dotted line (blue), 
and solid line (black) show the mass fall-back rates 
measured at $t=0$, $t=0.37$, $t=0.79$, and $t=4$, respectively. 
Each mass fallback rate is evaluated by using each differential 
mass distribution represented in panel~(a). The mass fallback rate and the orbital 
period are normalized by 1 solar mass per year and a year, respectively.
}
\label{fig:model1a}
\end{figure*}

\begin{figure*}
\resizebox{\hsize}{!}{
\includegraphics*[width=3mm]{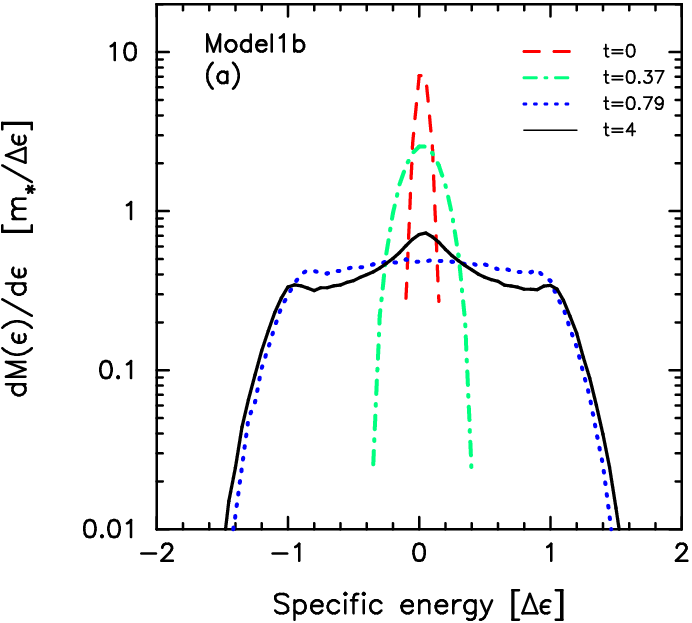}
\includegraphics*[width=3mm]{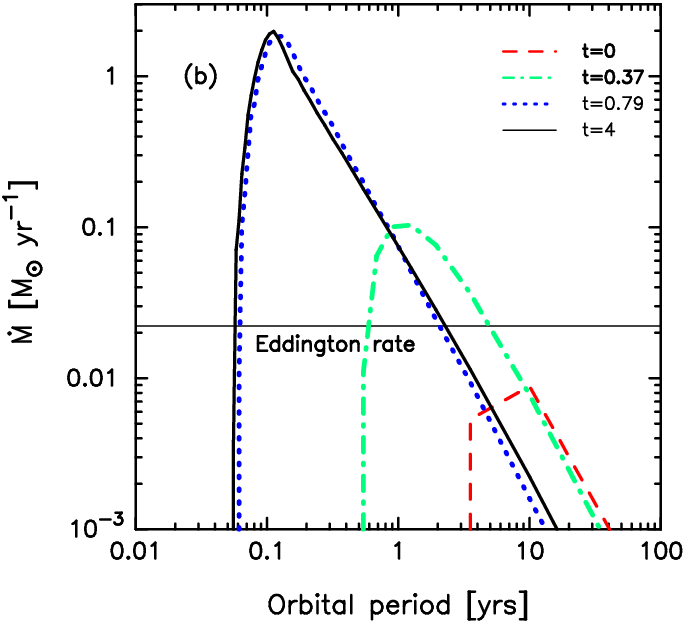}
}
\caption{Same as Fig.~\ref{fig:model1a}, but for Model~1b.}
\label{fig:model1b}
\end{figure*}

\begin{figure*}
\resizebox{\hsize}{!}{
\includegraphics*[width=3mm]{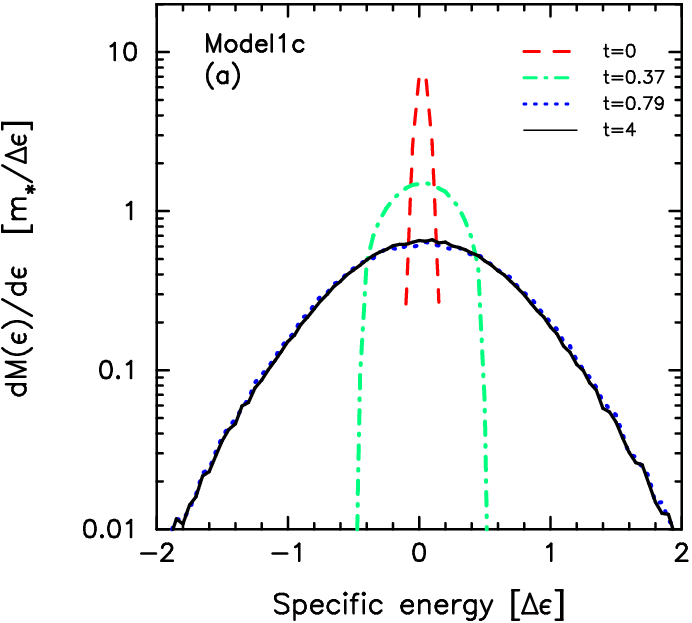}
\includegraphics*[width=3mm]{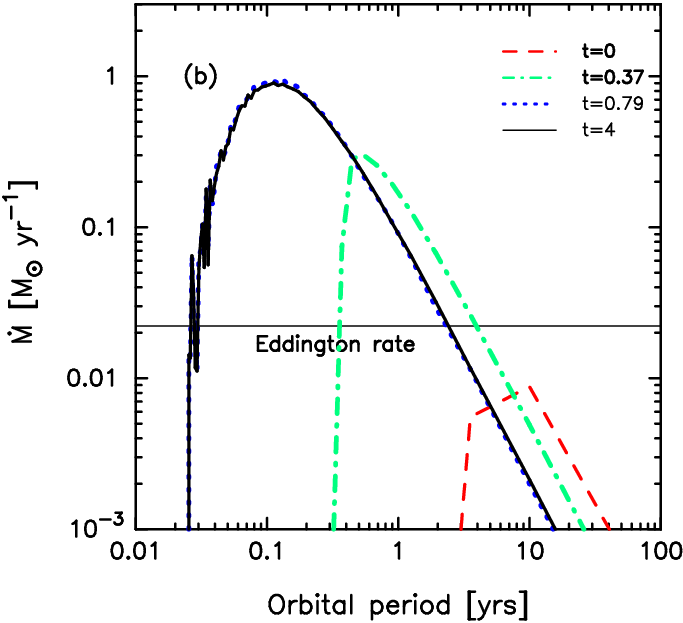}
}
\caption{Same as Fig.~\ref{fig:model1a}, but for Model~1c.}
\label{fig:model1c}
\end{figure*}

\begin{figure*}
\resizebox{\hsize}{!}{
\includegraphics*[width=3mm]{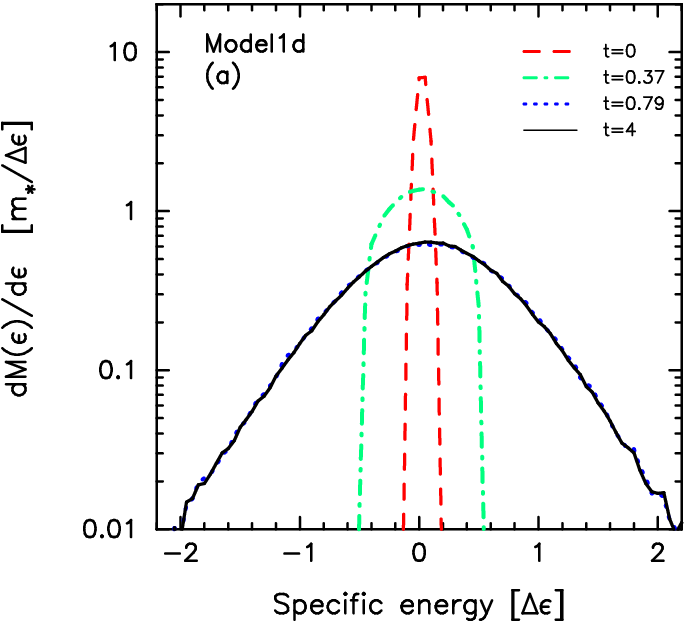}
\includegraphics*[width=3mm]{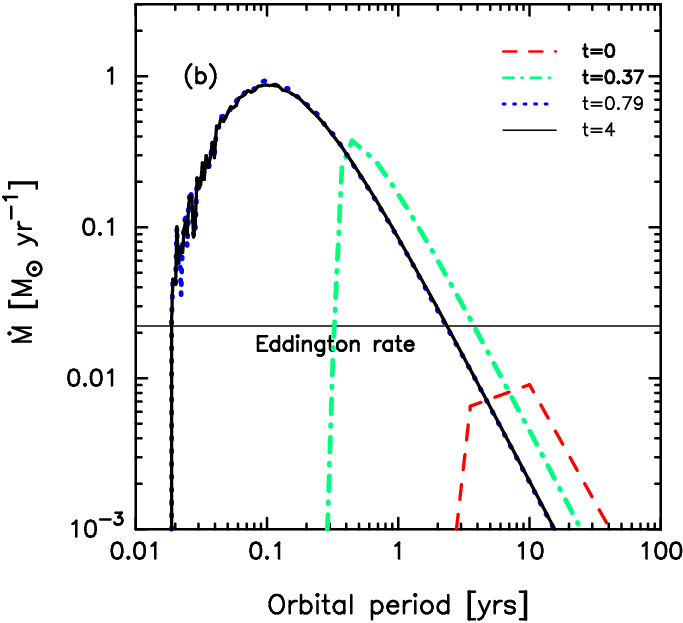}
}
\caption{Same as Fig.~\ref{fig:model1a}, but for Model~1d.}
\label{fig:model1d}
\end{figure*}

Figs.~\ref{fig:model1a}-\ref{fig:model1d} show the evolution 
of differential mass distributions and their corresponding 
mass fallback rates in Models~1a-1d. While the differential 
mass distribution is shown in panel (a), the mass fallback rate
is shown in panel (b).  In both panels, the solid line, dashed line, 
dot-dashed line, and dotted line represent the mass distributions and 
 corresponding mass fallback rates at $t=0.0$, $t=0.37$, $t=0.79$, 
 and $t=4.0$, respectively. Note that $t=0.5$ corresponds to the time 
 when the star reaches pericenter.
 In panel~(b), the horizontal solid line denotes the Eddington rate:
\begin{eqnarray}
\dot{M}_{\rm{Edd}}
&=&\frac{1}{\eta}\frac{L_{\rm{Edd}}}{c^2}
\simeq
2.2\times10^{-2}
\nonumber \\
&\times&
\left(\frac{\eta}{0.1}\right)^{-1}
\left(\frac{\it M_{\rm{BH}}}{10^6M_\odot}\right)\,\,M_\odot\rm{yr^{-1}},
\label{eq:eddrate}
\end{eqnarray}
where $L_{\rm{Edd}}=4\pi GM_{\rm{BH}}m_{\rm{p}}c/\sigma_{\rm{T}}$ 
is the Eddington luminosity with $m_{\rm{p}}$ 
and $\sigma_{\rm{T}}$ denoting the proton mass and 
Thomson scattering cross section, respectively, 
and $\eta$ is the mass-to-energy conversion efficiency,
which is set to $0.1$ in the following discussion.

In Model~1a, the mass distribution broadens with time as
the stellar debris is put on near-ballistic orbits by tidal interaction with the black hole.
In panel~(a), the central peak formed after $t\ga1.0$ is due 
to mass congregation, from the self-gravity of the stellar debris (in this scenario, the star is barely disrupted).
The energy spread corresponds to $\Delta\epsilon$ before and 
after the tidal disruption. The corresponding mass fallback rates 
are proportional to $t^{-5/3}$ except for the solid line in panel (b). 
This is in good agreement with the literatures \citep{rmj88, ecr89}.
The slight deviation from t to the $-5/3$ power originates from the 
convexity around $\Delta\epsilon$ due to re-congregation of stellar mass under self-gravity (see \citet{lg09}).

Model~1b has the same simulation parameters as Model~1a 
except that it adopts the pseudo-Newtonian potential. In panel~(a) 
of Fig.~\ref{fig:model1b}, the final energy spread and the central 
peak are slightly wider and milder than those of Model~1a. This is 
because the energy imparted by the tidal force under the pseudo-Newtonian
potential is slightly larger than that of the Newtonian potential.
From equation~(\ref{eq:wegg}), the energy spread under the 
pseudo-Newtonian potential is evaluated to be
\begin{equation}
\Delta\epsilon_{\rm{PN}}\approx\frac{dU(r)}{dr}\biggr|_{r=r_{\rm{t}}}r_{*}
=\Delta\epsilon\left[c_1+\frac{1-c_1}{[1-c_2(r_{\rm{S}}/2r_{\rm{t}})]^2}+
2c_3\frac{r_{\rm{S}}}{2r_{\rm{t}}}\right].
\label{eq:deltaew}
\end{equation}
The expected energy spread $\Delta\epsilon_{\rm{PN}}\sim1.1\Delta\epsilon$ 
for $10^6M_\odot$ is in good agreement with the simulated energy spread.

Model~1c has the same simulation parameters as Model~1a 
except for $\beta=5$, while Model~1d has the same simulation 
parameters as Model~1c except for adopting the pseudo-Newtonian 
potential. For higher value of $\beta$ than unity, the tidal forces acting 
on the star become stronger because the pericenter distance is smaller 
than the tidal disruption radius. According to equation~(2) of \citet{ecr89}, 
the energy spread of the stellar debris can be estimated to be 
$\beta^2\Delta\epsilon$, and should become $25\Delta\epsilon$ for Models 1c and 1d.
In these two models, however, the energy spread is only about $1.4$ times wider 
than that of standard model, showing that the energy spread weakly depends on $\beta$. 
This is in agreement with recent numerical \citep{grr12} and analytic work \citep{ssl12} 
that demonstrates the inaccuracy of the traditional formula for $\Delta\epsilon$, 
in place of which we have used the more accurate equation~(\ref{eq:spreade}).
Since the tidal disruption is less marginal as $\beta$ is higher, the re-congregation of the mass 
due to self-gravity of the stellar debris is prevented. This leads to the mildly-sloped 
mass distribution, and therefore the peak of the mass fallback rate also smooths.
There is no remarkable difference between Model~1c and Model~1d except
that the energy spread of Model~1d is slightly wider than that of Model~1c 
following equation~(\ref{eq:deltaew}).

%
%
%
%
%
%
%
%
%
%
%
%
\subsection{Eccentric tidal disruption}
\label{sec:ecctde}
%
The specific orbital energy of a star on an eccentric orbit is given by
\begin{equation}
\epsilon_{\rm{orb}}\approx-\frac{GM_{\rm{BH}}}{2a_*}
=-\frac{GM_{\rm{BH}}}{2r_{\rm{t}}}\beta(1-e_*).
\label{eq:eporb}
\end{equation}
This quantity is less than zero because of the finite
value of $a_*$, in contrast to the standard, parabolic orbit of a
star.  If $\epsilon_{\rm{orb}}$ is less than $\Delta\epsilon$, all the
stellar debris should be bounded to the black hole, even after the
tidal disruption. The condition $\epsilon_{\rm{orb}}=\Delta\epsilon$
therefore gives a critical value of orbital eccentricity of the star
\begin{equation}
e_{\rm crit}\approx1-\frac{2}{\beta}\left( \frac{m_*}{M_{\rm BH}} \right)^{1/3},
\label{eq:ecrit}
\end{equation}
below which all the stellar debris should remain gravitationally 
bound to the black hole  \citep{asp11}. The critical eccentricity is evaluated to be 
$e_{\rm{crit}}=0.996$ for Model~2a and Model~3b, 
whereas $e_{\rm{crit}}=0.98$ for Model~3a.

For the eccentric TDEs, $t_{\rm{min}}$ and $t_{\rm{max}}$ are obtained 
by substituting $\epsilon=\Delta\epsilon\pm \epsilon_{\rm{orb}}$ into 
equation~(\ref{eq:te}). The orbital period of the most tightly bound orbit is given by
\begin{equation}
t_{\rm min} =\frac{\pi}{\sqrt{2}}\frac{\Omega_{*}^{-1}}{\beta^{3/2}(1-e_{*})^{3/2}}.
\label{eq:tmin}
\end{equation}
The orbital period of the most loosely bound orbit $t_{\rm{max}}$ is estimated to 
be $\infty$ for $e_*\ge{e_{\rm{crit}}}$, while ${t_{\rm{max}}}$ converges with
\begin{equation}
t_{\rm max}=\frac{\pi}{\sqrt{2}}\Omega_{*}^{-1}
\left[ 
\frac{\beta(1-e_*)}{2} -
\left(
\frac{m_*}{M_{\rm{BH}}}
\right)^{1/3} 
\right]^{-3/2}
\label{eq:tmax}
\end{equation}
for $e_*<e_{\rm crit}$.
The duration time of mass fallback for eccentric TDEs with $e_*<e_{\rm{crit}}$ 
is thus predicted to be finite and can be written by 
\begin{eqnarray}
&&
\Delta{t}
=t_{\rm{max}} - t_{\rm{min}}
=\frac{\pi}{\sqrt{2}}
\frac{\Omega_{*}^{-1}}{
[\beta(1-e_{*})]^{3/2}}
\nonumber \\
&&
\times
\left(
\left[ 
\frac{1}{2} -
\frac{1}{\beta(1-e_{*})}
\left(
\frac{m_*}{M_{\rm{BH}}}
\right)^{1/3} 
\right]^{-3/2}
-1
\right).
\label{eq:deltat}
\end{eqnarray}
Evaluating this gives $\Delta{t}\approx4.3\Omega_*^{-1}$ for Model~2a and 
$\Delta{t}\approx207\Omega_{*}^{-1}$ for Model~3b, whereas 
$\Delta{t}\rightarrow\infty$ for Model~3a.

\begin{figure*}
\resizebox{\hsize}{!}{
\includegraphics*[width=3mm]{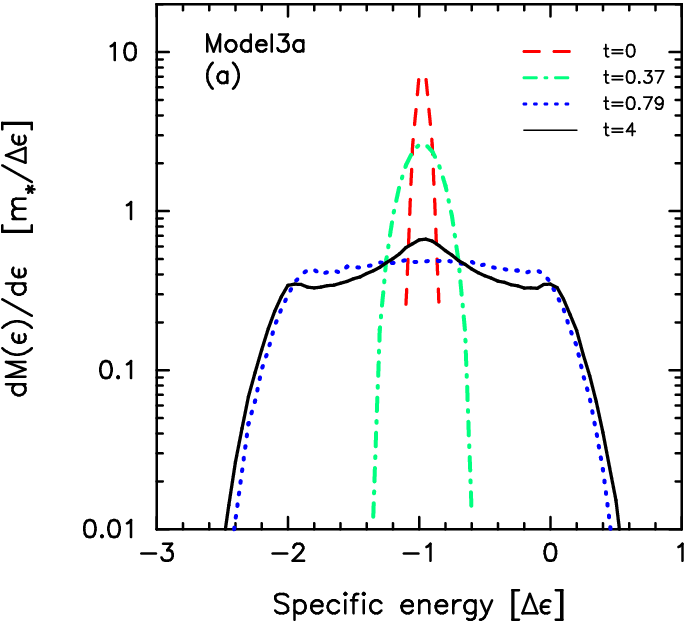}
\includegraphics*[width=3mm]{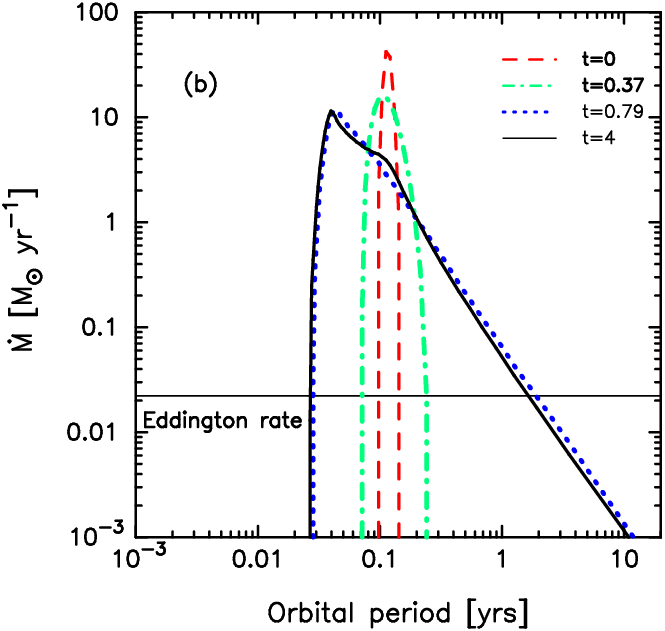}
}
\caption{Same as Fig.~\ref{fig:model1a}, but for Model~3a.}
\label{fig:model3a}
\end{figure*}

\begin{figure*}
\resizebox{\hsize}{!}{
\includegraphics*[width=3mm]{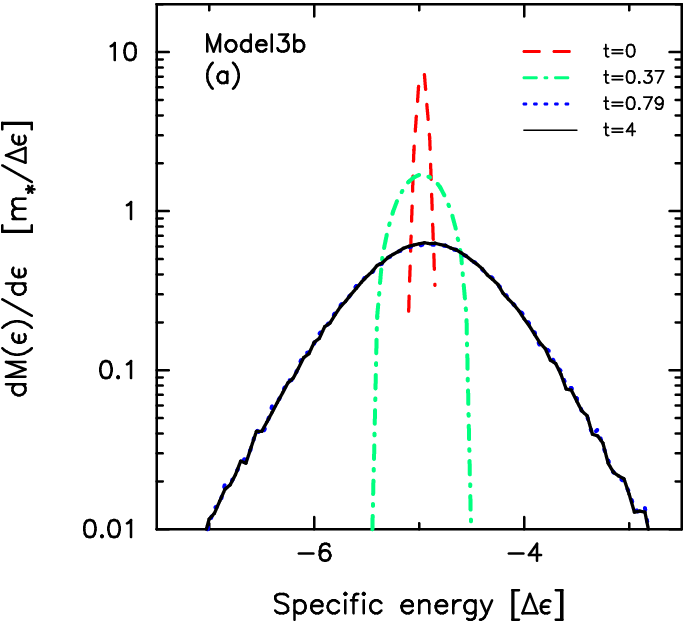}
\includegraphics*[width=3mm]{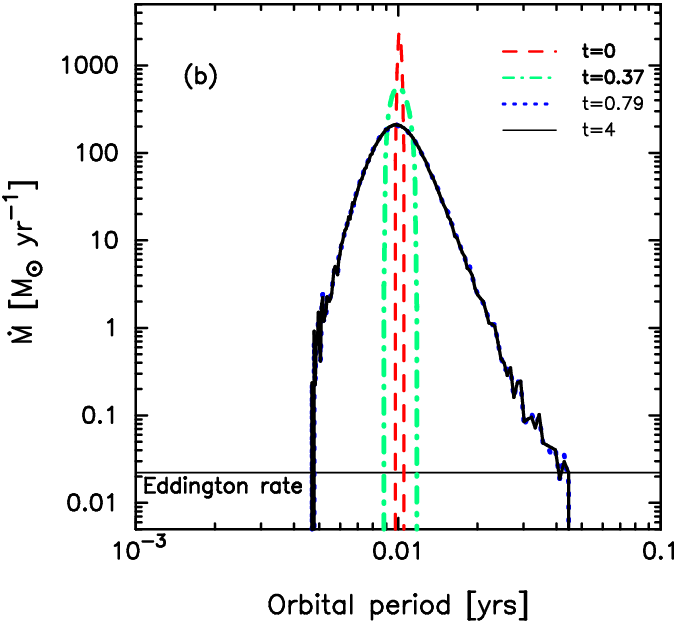}
}
\caption{Same as Fig.~\ref{fig:model1a}, but for Model~3b.}
\label{fig:model3b}
\end{figure*}

Figs.~\ref{fig:model3a} and \ref{fig:model3b} show the evolution of
differential mass distributions and corresponding mass fallback rates
in Model~3a and Model~3b, respectively.  The figure formats are the
same as Figs.~\ref{fig:model1a}-\ref{fig:model1d}.  The mass is not
distributed around zero but around $-\Delta\epsilon$ in Model~3a, and
around $-5\Delta\epsilon$ in Model~3b.  This is because the specific
energy of the initial stellar orbit is originally negative (see
equation~\ref{eq:eporb}).  It is clear from the negative shift of the
mass distribution's center that most of the mass in Model~3a
and all of the mass in Model~3b are bounded.  As shown in panel (a) of Fig.~\ref{fig:model3a}, the
resultant energy spread is slightly larger than we analytically
expected.  This suggests that the critical eccentricity is slightly smaller
than the value in equation~(\ref{eq:ecrit}).

From equations~(\ref{eq:tmin}) and (\ref{eq:tmax}),
$t_{\rm{min}}=0.04\rm{yr}$ and ${t}_{\rm{max}}\rightarrow\infty$ for Model~3a, 
and $t_{\rm{min}}=0.0035\rm{yr}$ and $t_{\rm{max}}=0.014\rm{yr}$ for Model~3b.
These values deviate from simulation results, because they were derived assuming a 
smaller spread in energy than in our simulations.
Furthermore, $t_{\rm{min}}$ is shorter than that of the parabolic case.
This makes the mass fallback rate about one order of magnitude higher.
Notably, the mass accretion is completely finite in Model~3b 
and its rate is enhanced by about two orders of magnitude, because its timescale 
is much shorter than that of the standard model.

%
%
%
%
%
\subsection{Accretion disk formation}
\label{sec:diskform}
%
%
%
\begin{figure*}
\resizebox{\hsize}{!}{
\includegraphics*[width=3mm]{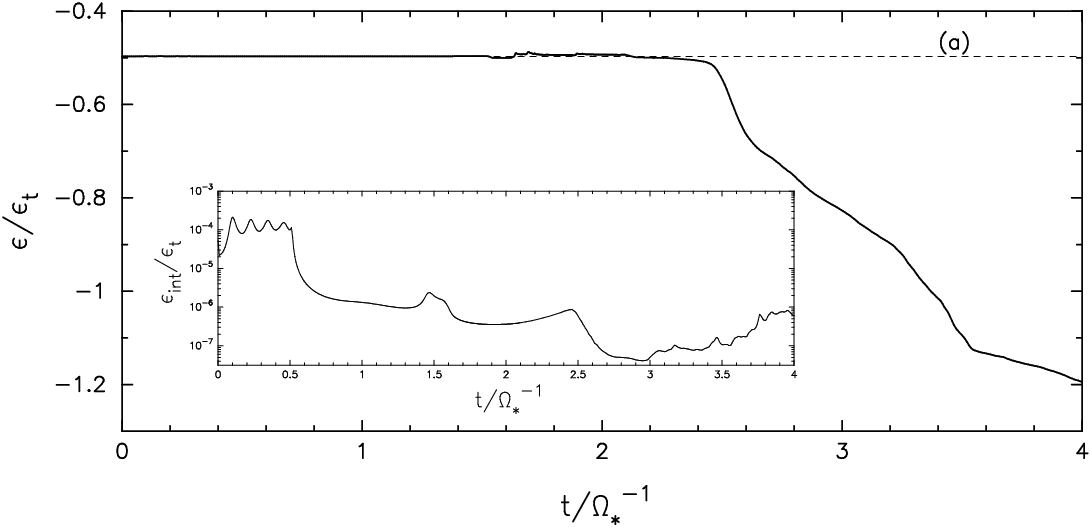}
}\\
\resizebox{\hsize}{!}{
\includegraphics*[width=3mm]{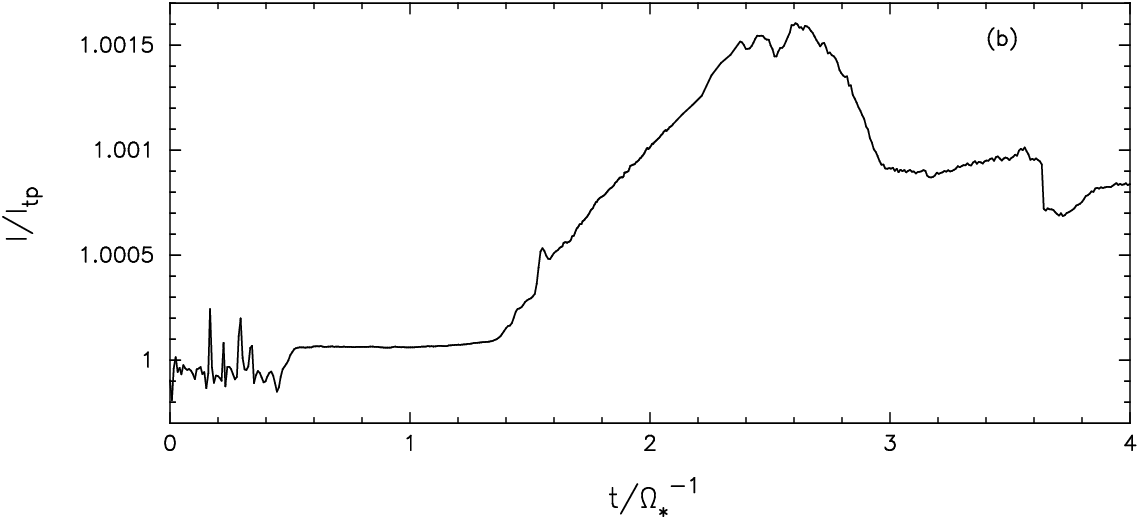}
}
\caption{
Evolution of the specific energy and specific angular momentum in Model~2a.
In panel (a), the solid line and dashed line represent the specific energy of SPH particles 
and that of a test particle $\epsilon_{\rm{tp}}=-0.497$, respectively. 
The small inside panel shows the evolution of the specific internal energy, $\epsilon_{\rm{int}}$.
They are normalized by $\epsilon_{\rm{t}}=GM/r_{\rm{t}}$.
Panel (b) shows the specific angular momentum  normalized by $l_{\rm{tp}}=0.714$,
which is the specific angular momentum of the initial stellar orbit.
}
\label{fig:eevol} 
\end{figure*}

%
\begin{figure*}
\resizebox{\hsize}{!}{
\includegraphics*[width=36mm]{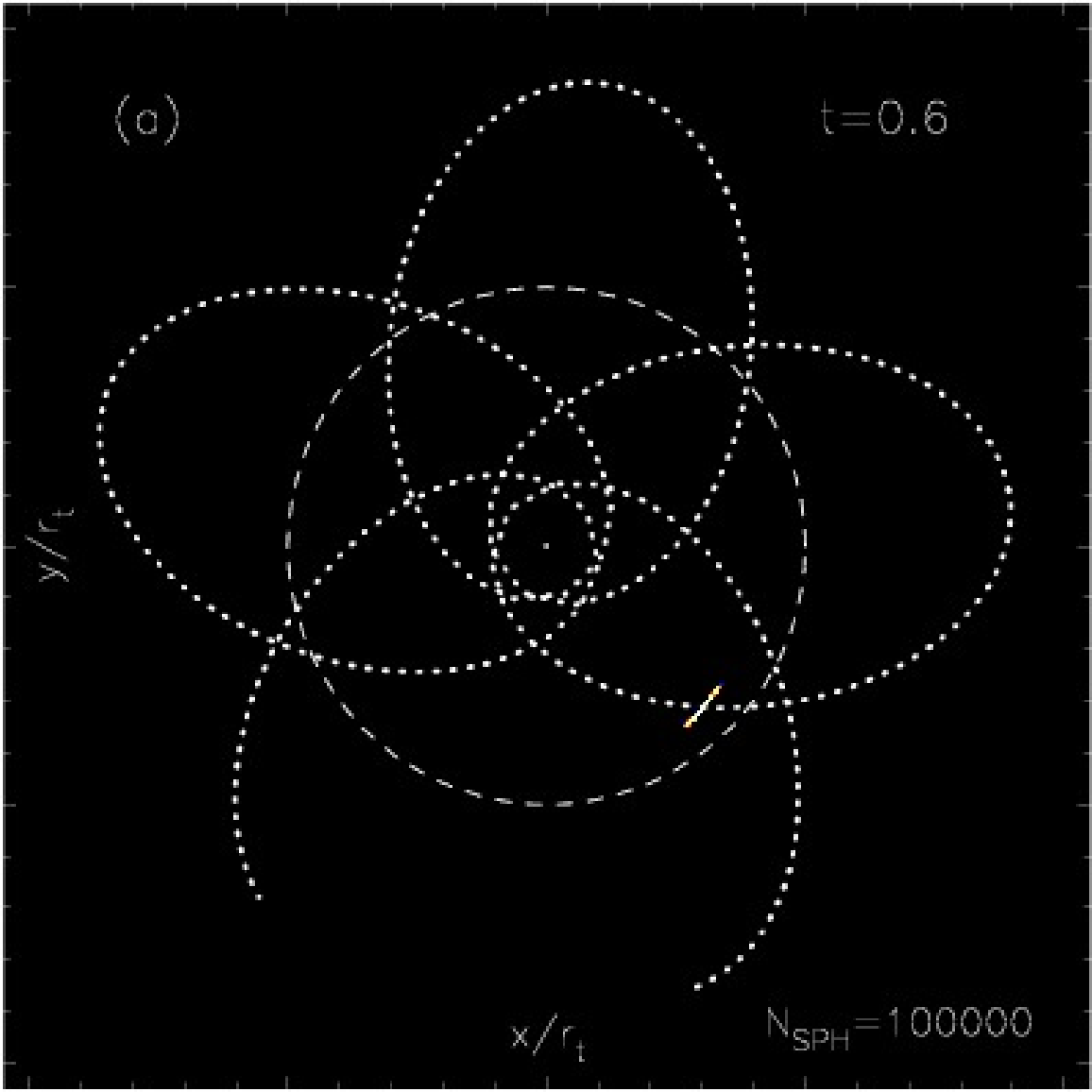}
\includegraphics*[width=36mm]{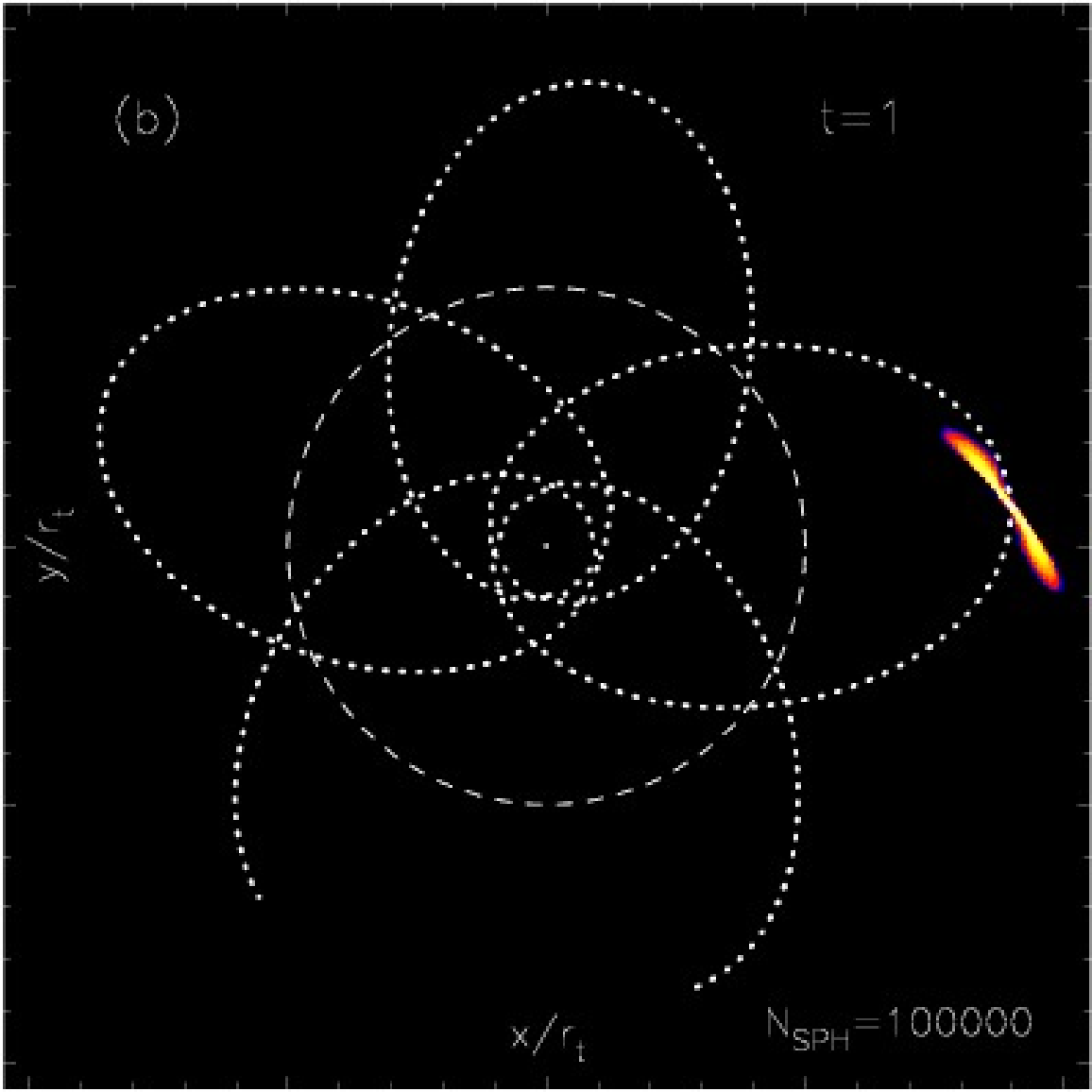}
\includegraphics*[width=36mm]{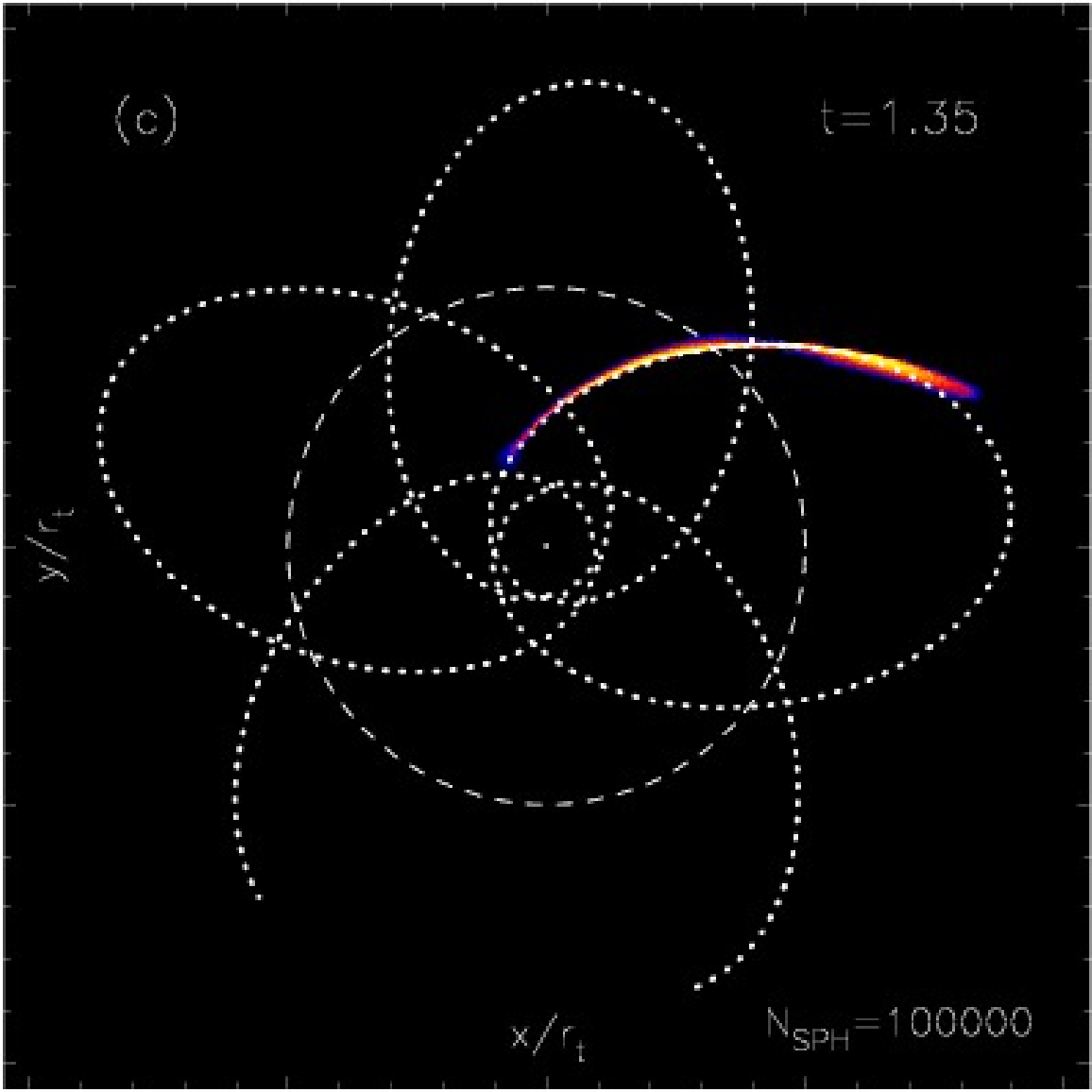}
}\\
\resizebox{\hsize}{!}{
\includegraphics*[width=36mm]{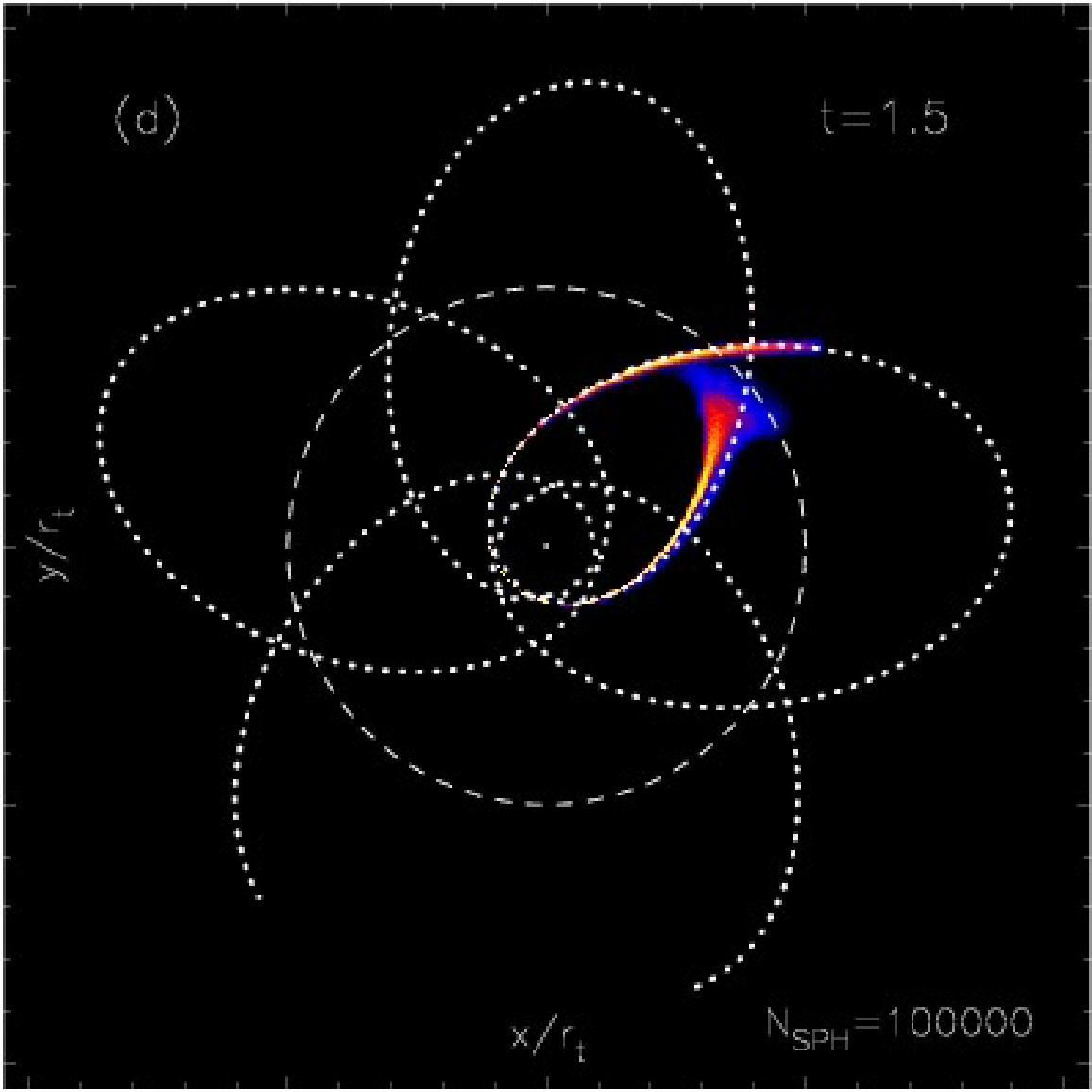}
\includegraphics*[width=36mm]{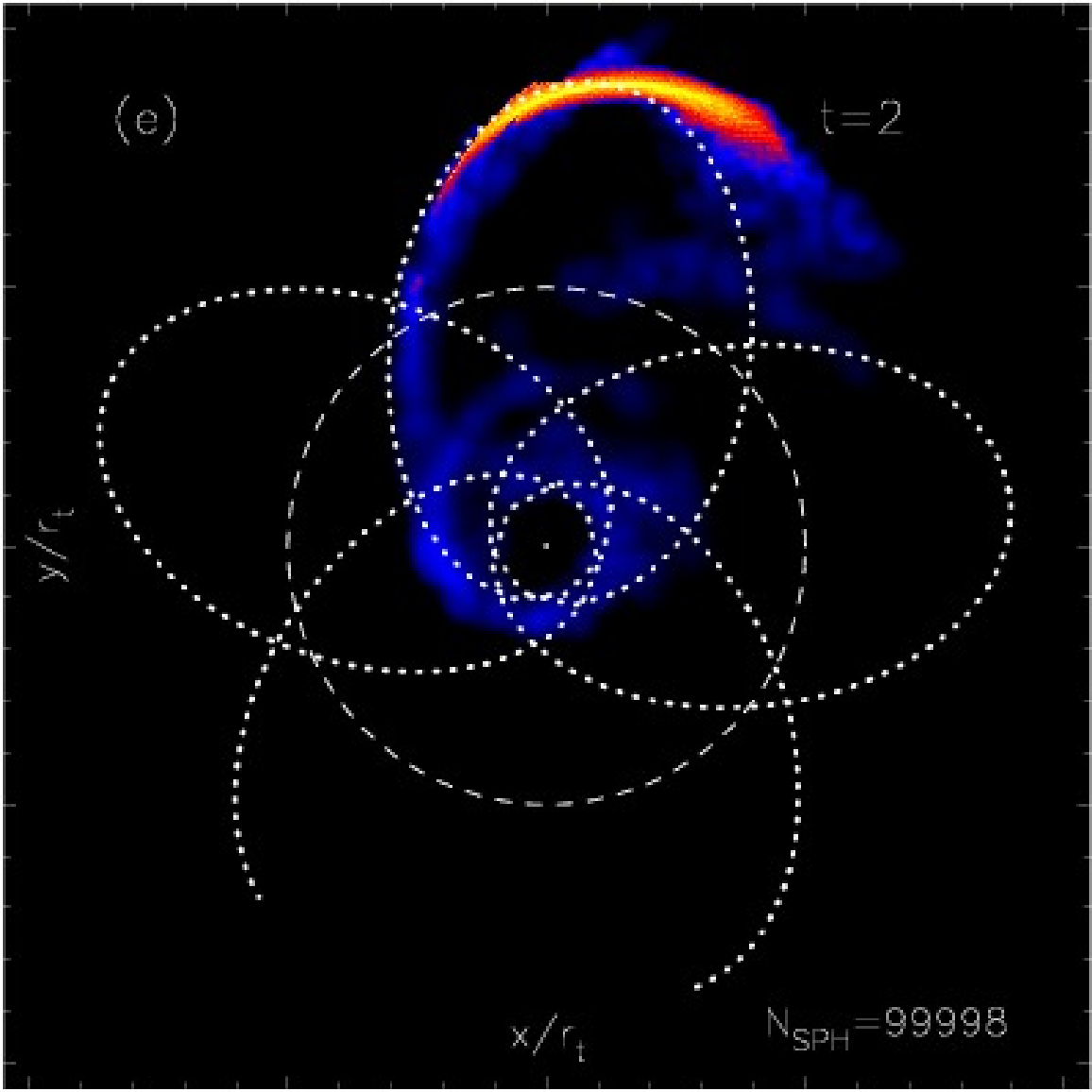}
\includegraphics*[width=36mm]{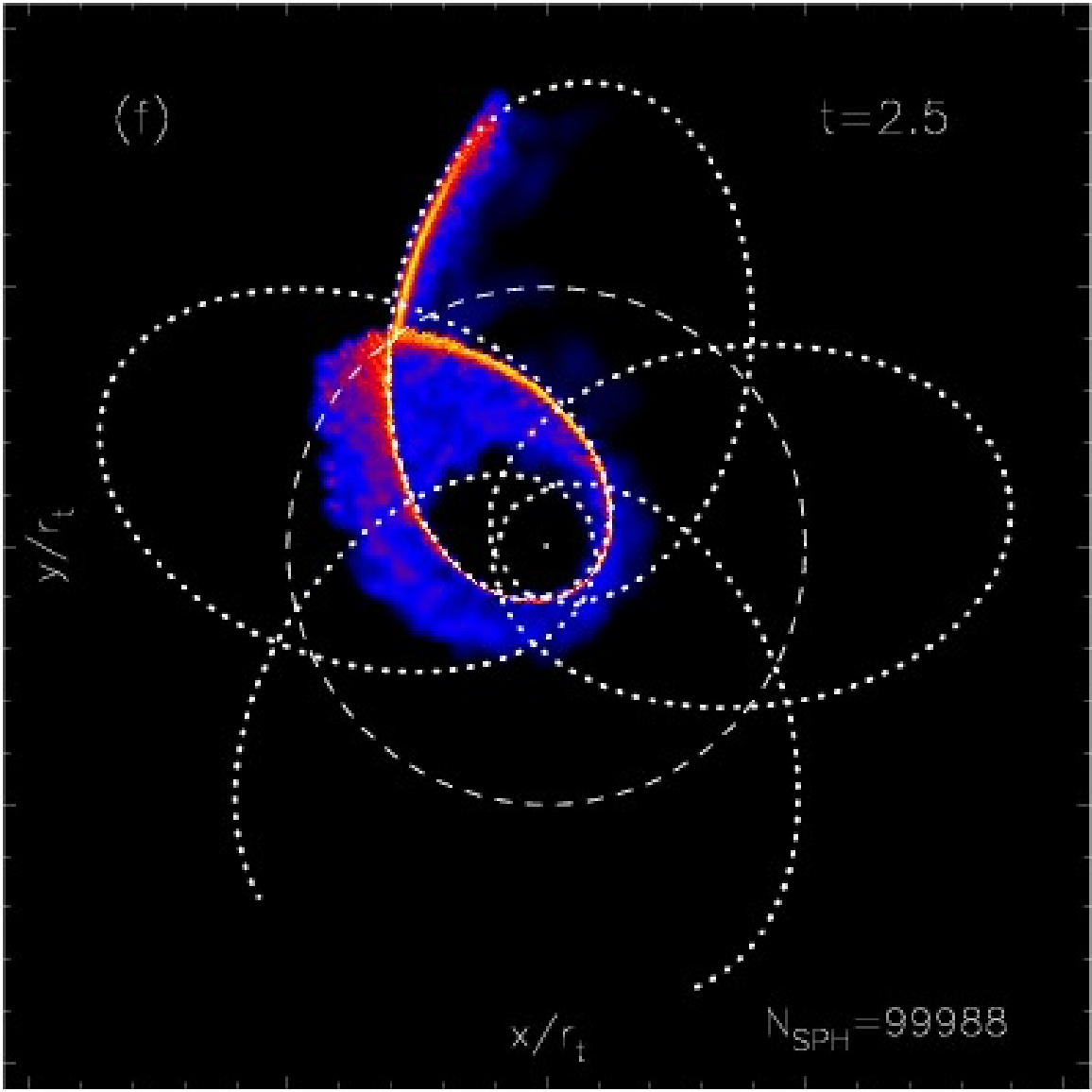}
}\\
\resizebox{\hsize}{!}{
\includegraphics*[width=36mm]{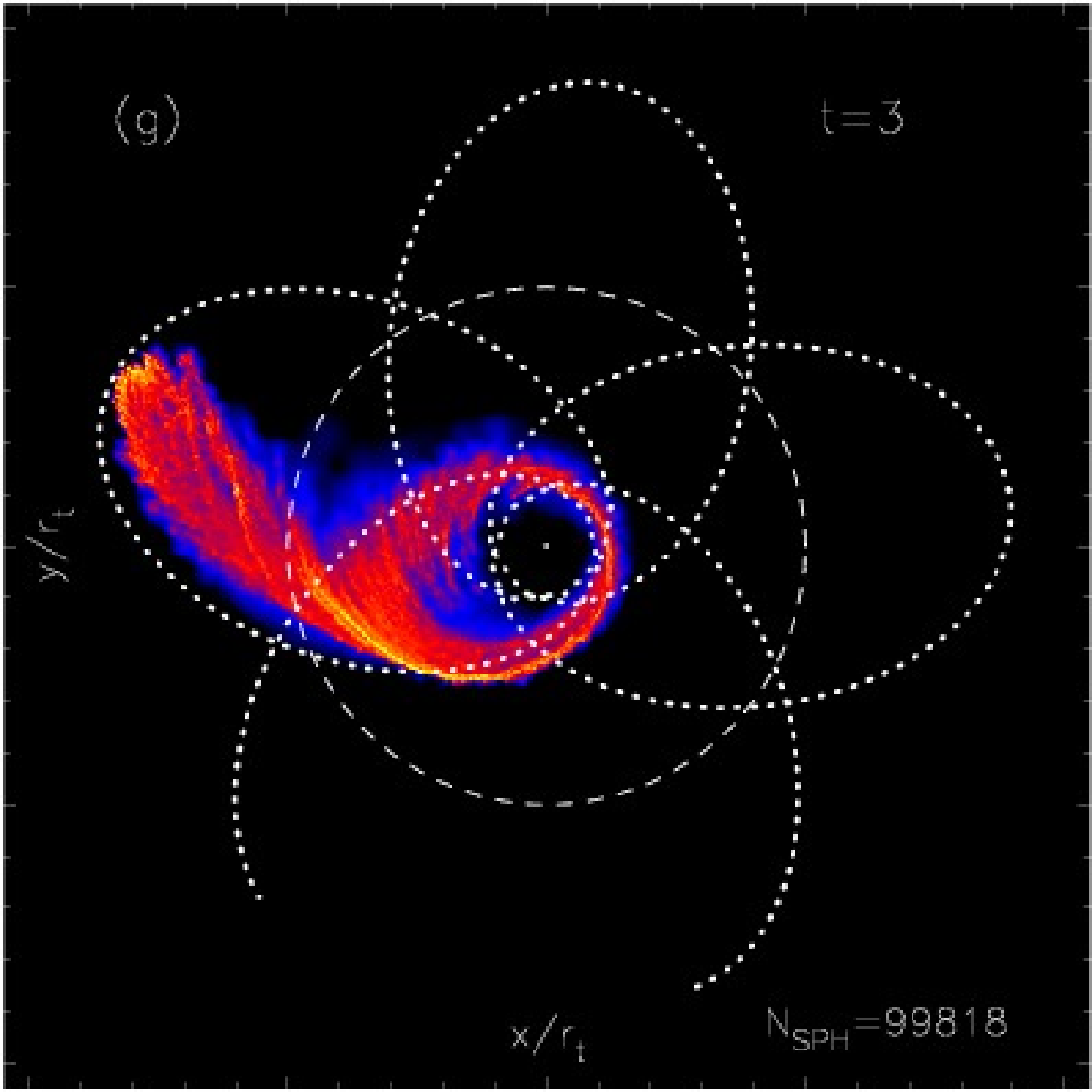}
\includegraphics*[width=36mm]{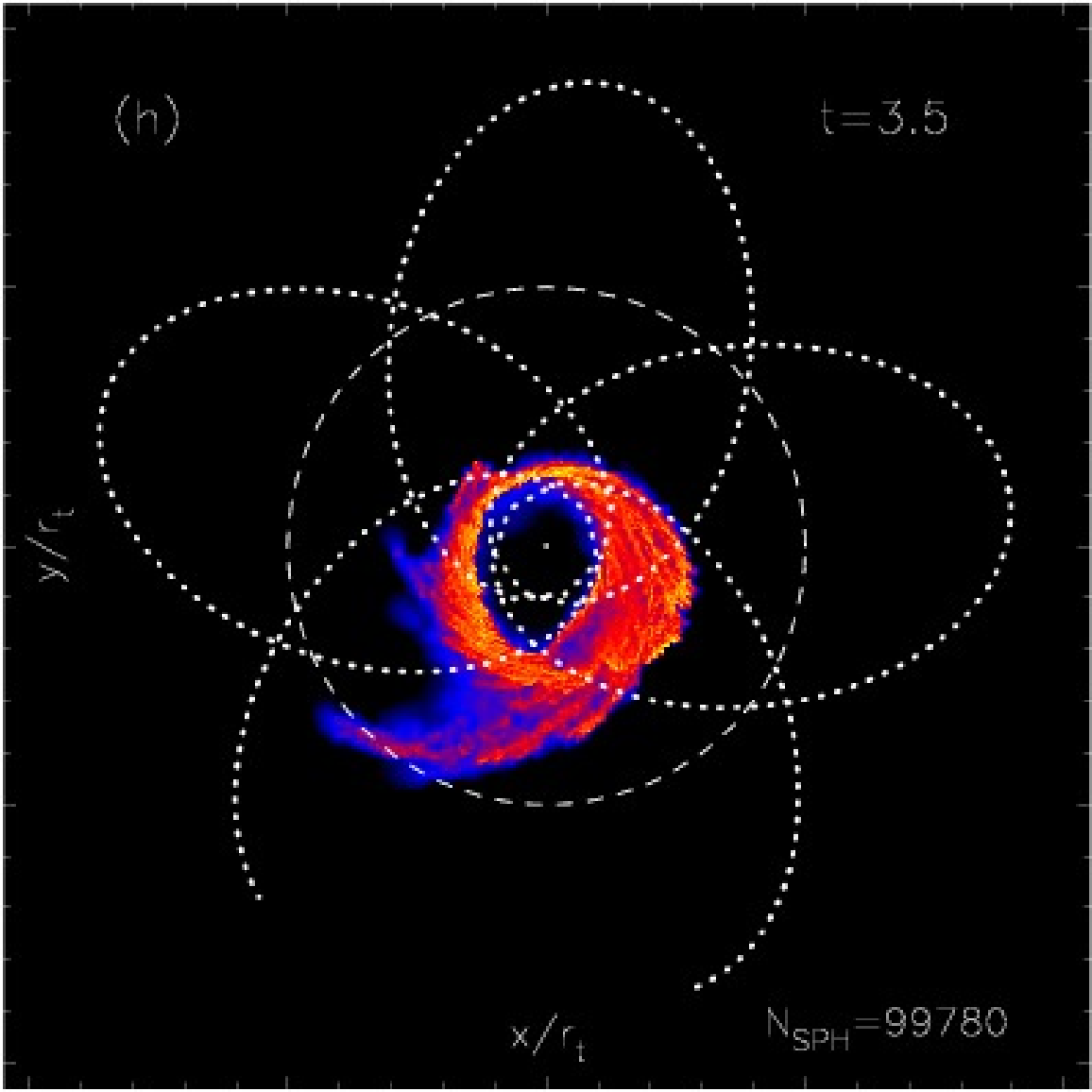}
\includegraphics*[width=36mm]{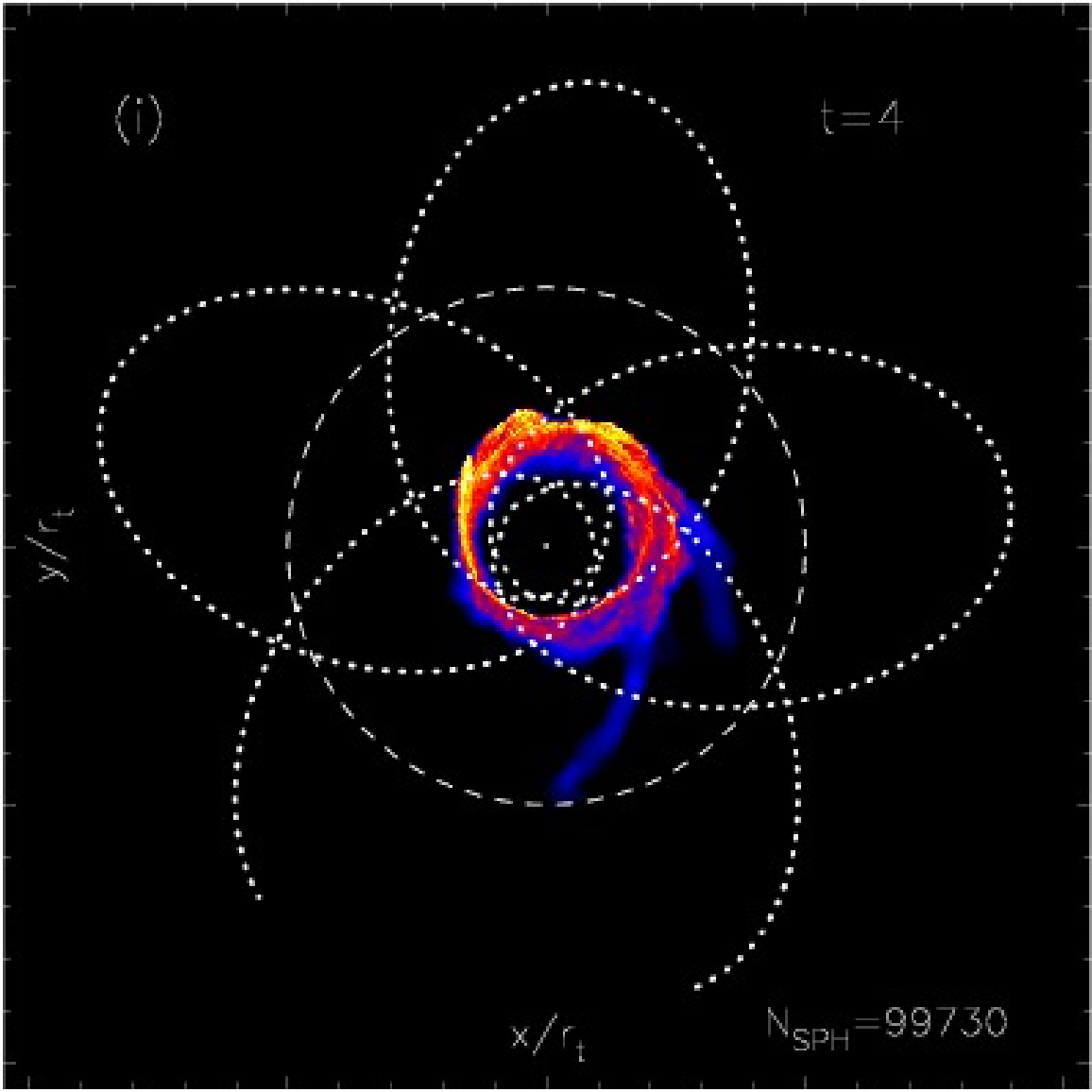}
}
\caption{
A sequence of snapshots of the tidal disruption process in Model~2a. 
They are from panel (a) to panel (i) in chronological order. 
Each panel shows a surface density projected on $x$-$y$ plane 
in five orders of magnitude in a logarithmic scale for $0.6\le{t}\le4$, 
where $t$ is in units of $\Omega_{*}^{-1}$. 
The black hole is set at the origin. The run time is annotated at the top-right corner, 
while the number of SPH particles are indicated at the bottom-right corner. 
The dashed circle and dotted line indicate the tidal disruption radius 
and the orbits of a test particle moving under the pseudo-Newtonian potential 
given by equation~(\ref{eq:wegg}), respectively.
}
\label{fig:diskform}
\end{figure*}
%
\begin{figure*}
\resizebox{\hsize}{!}{
\includegraphics*[width=36mm]{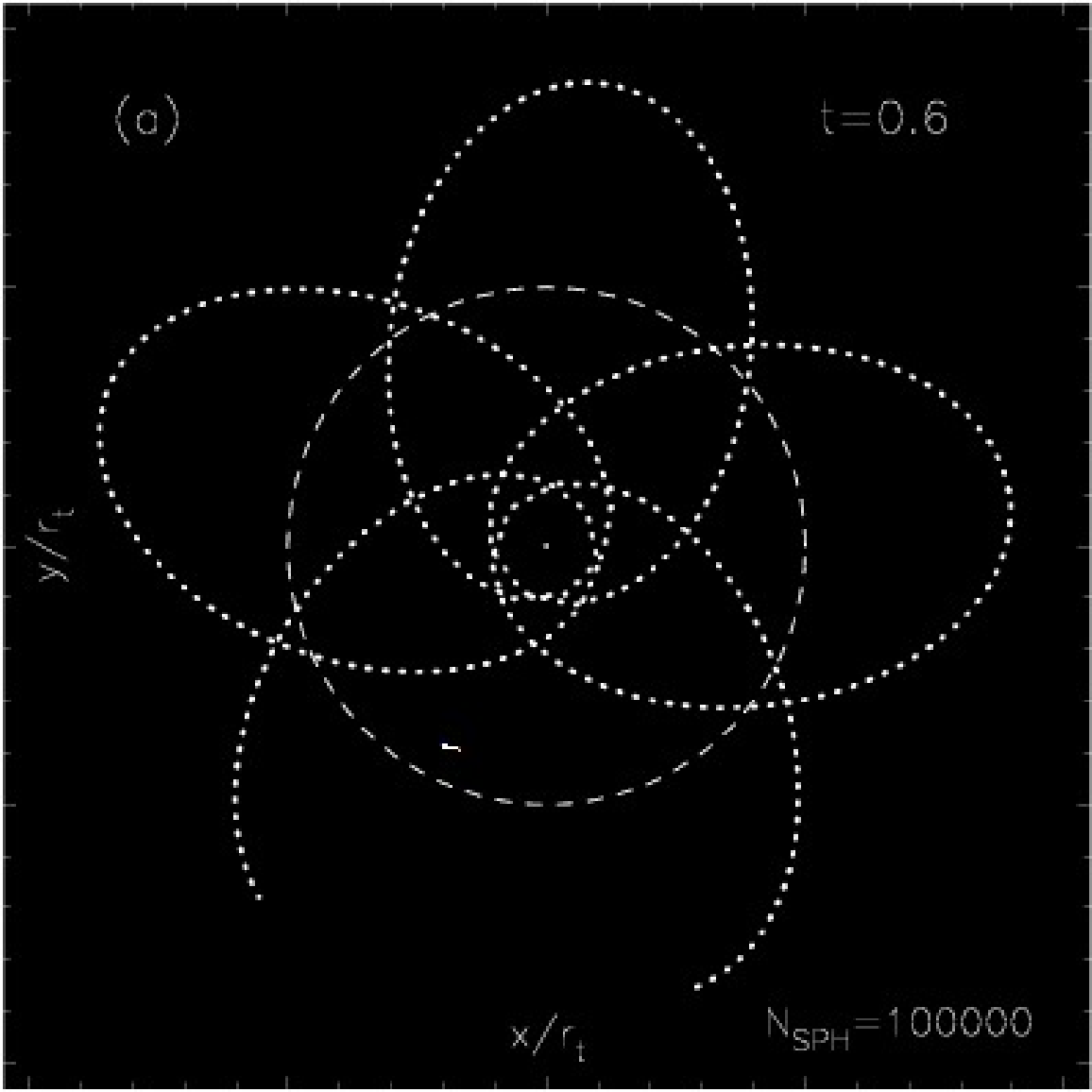}
\includegraphics*[width=36mm]{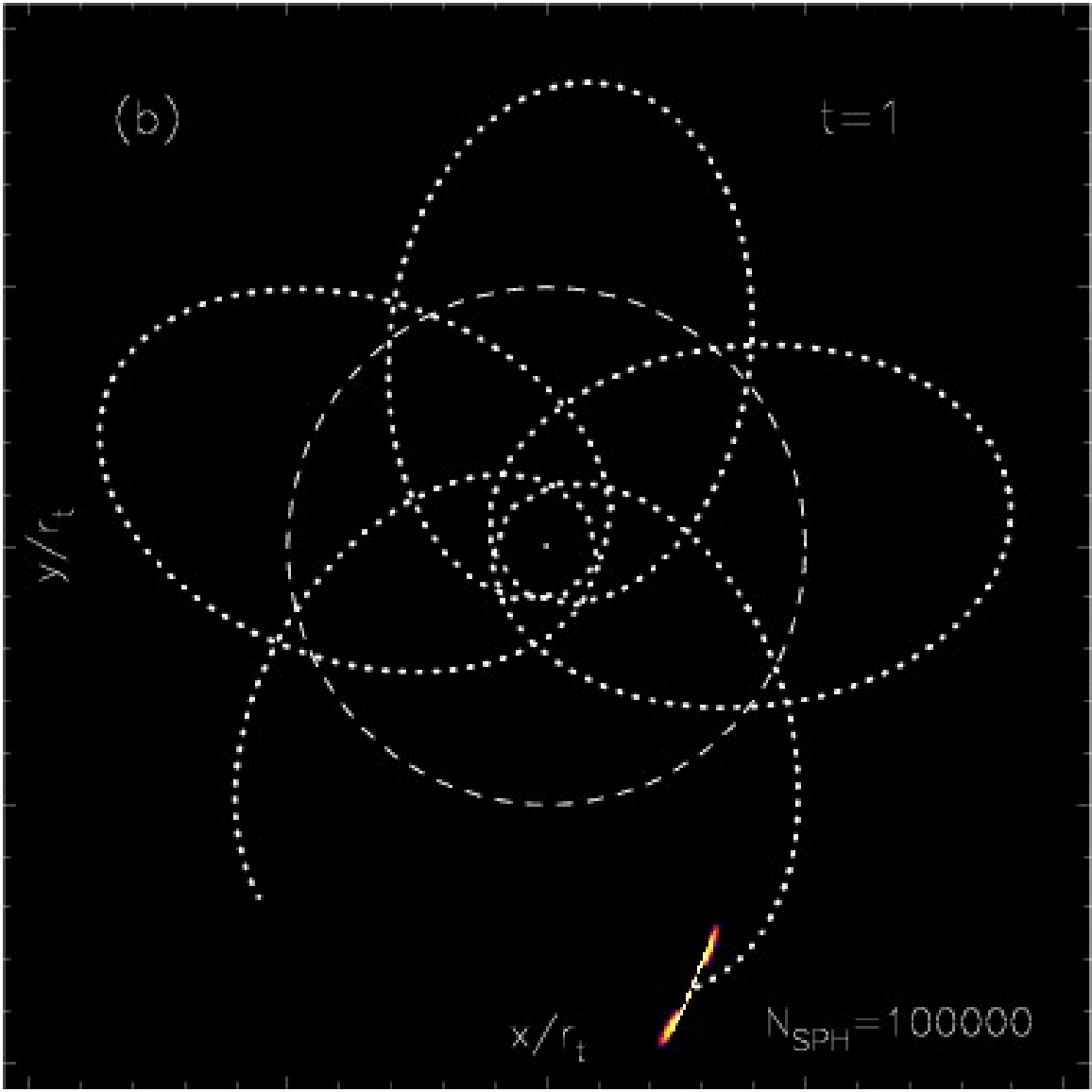}
\includegraphics*[width=36mm]{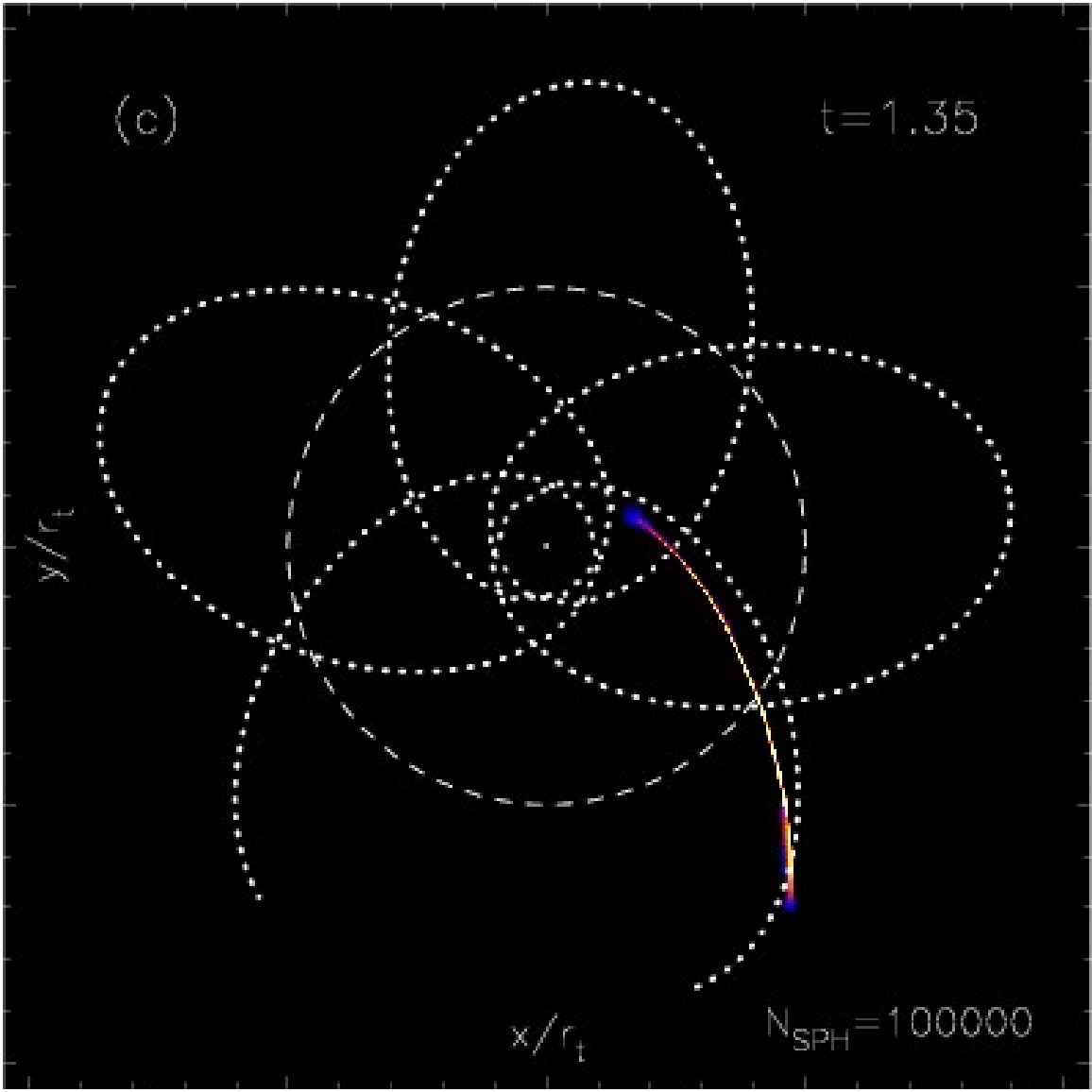}
}\\
\resizebox{\hsize}{!}{
\includegraphics*[width=36mm]{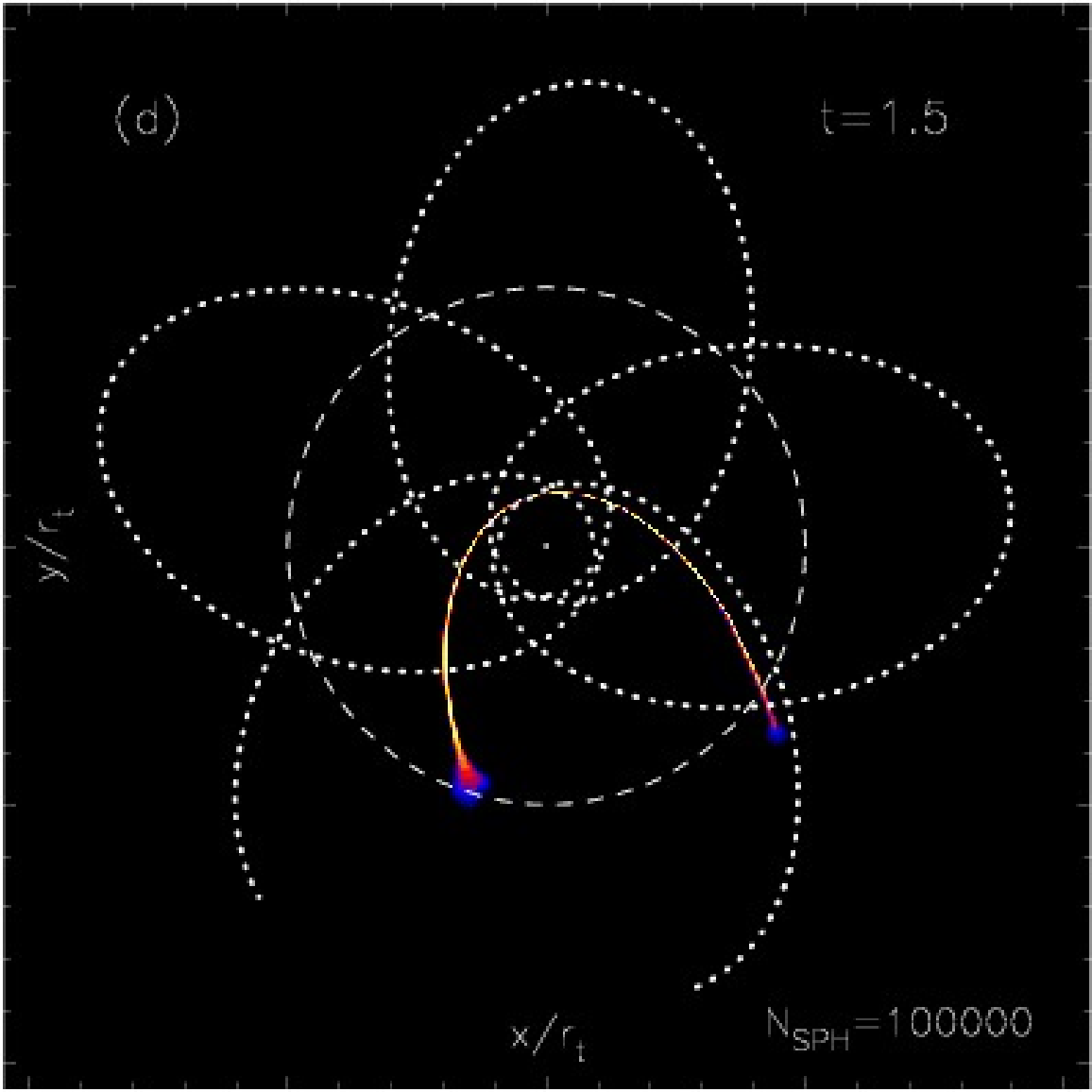}
\includegraphics*[width=36mm]{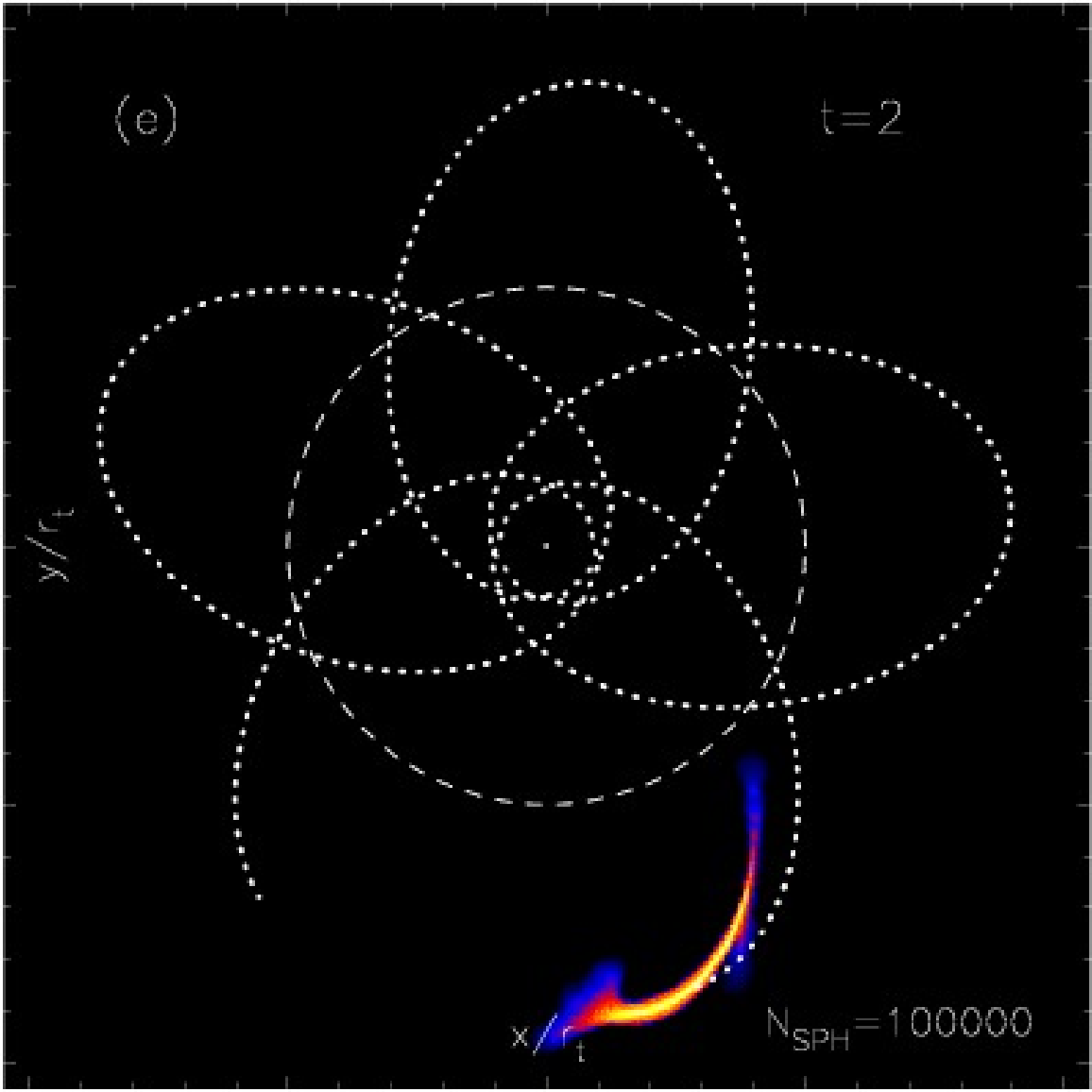}
\includegraphics*[width=36mm]{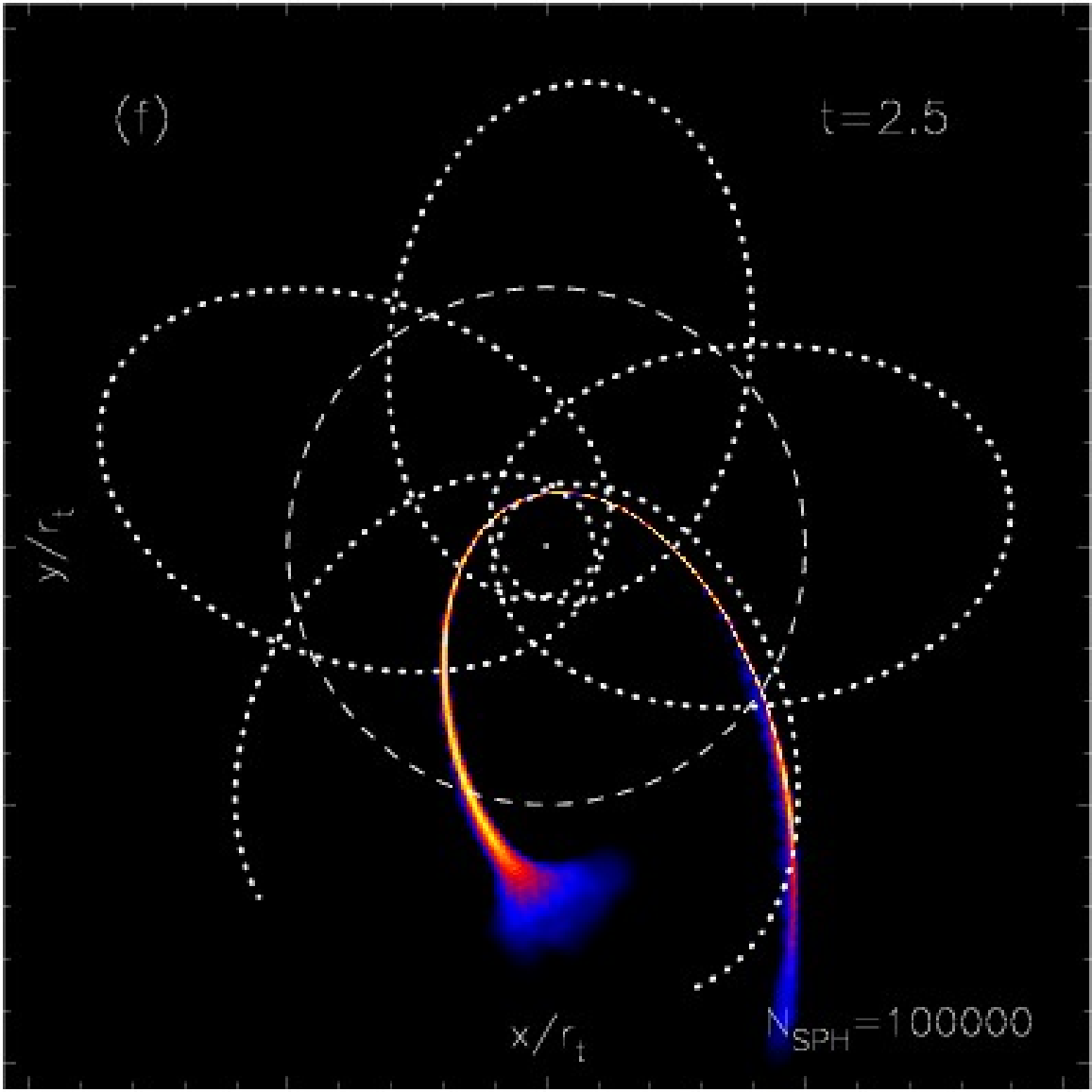}
}\\
\resizebox{\hsize}{!}{
\includegraphics*[width=36mm]{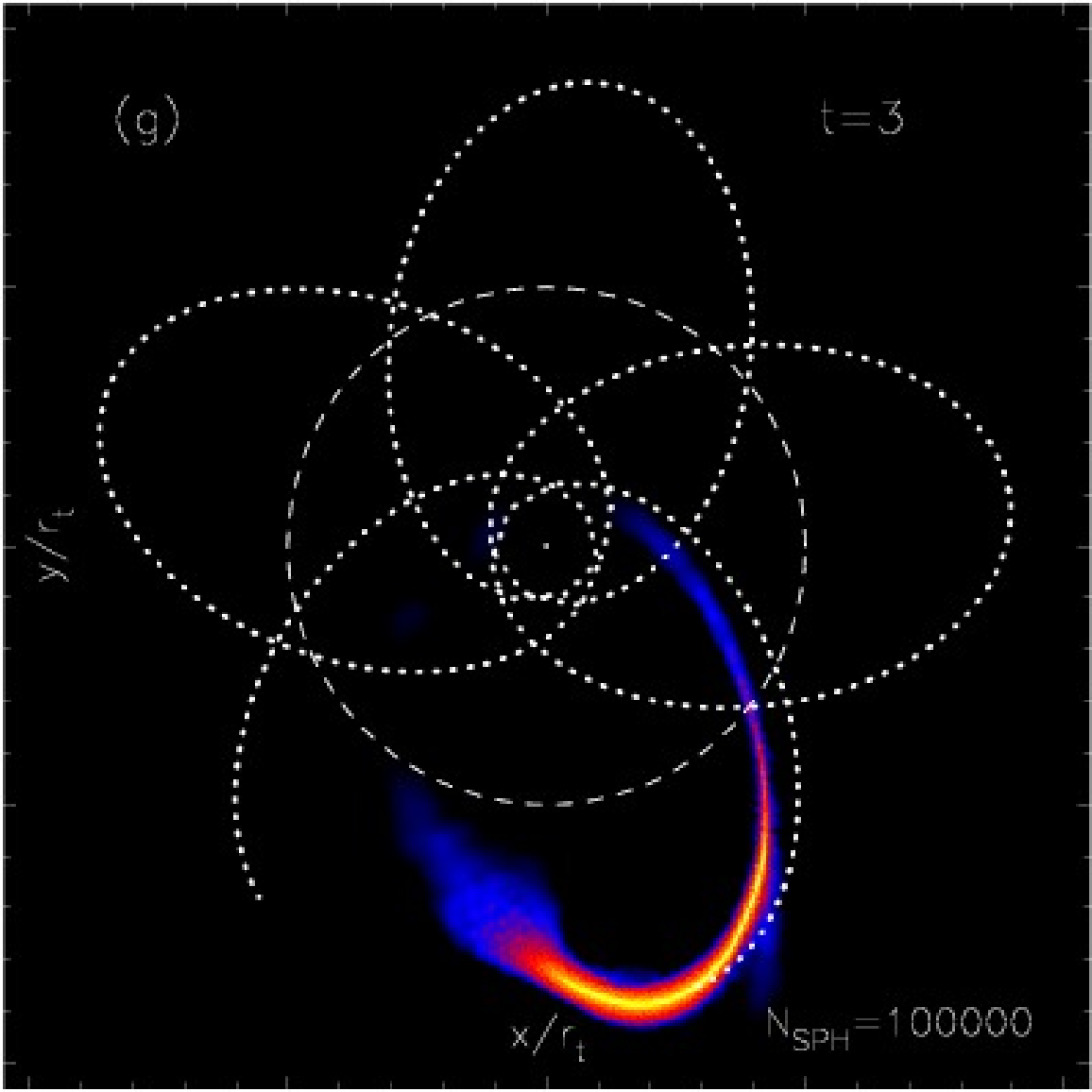}
\includegraphics*[width=36mm]{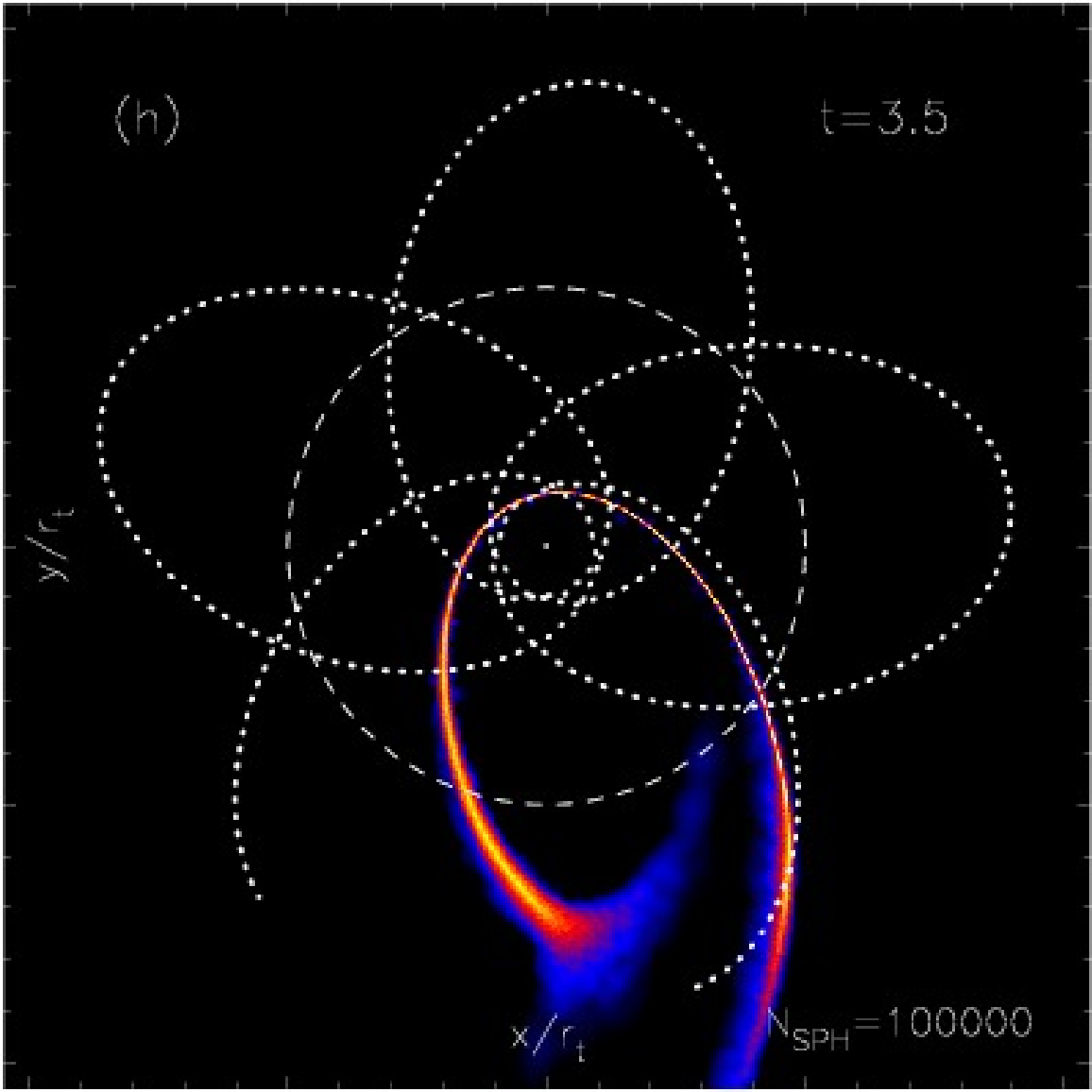}
\includegraphics*[width=36mm]{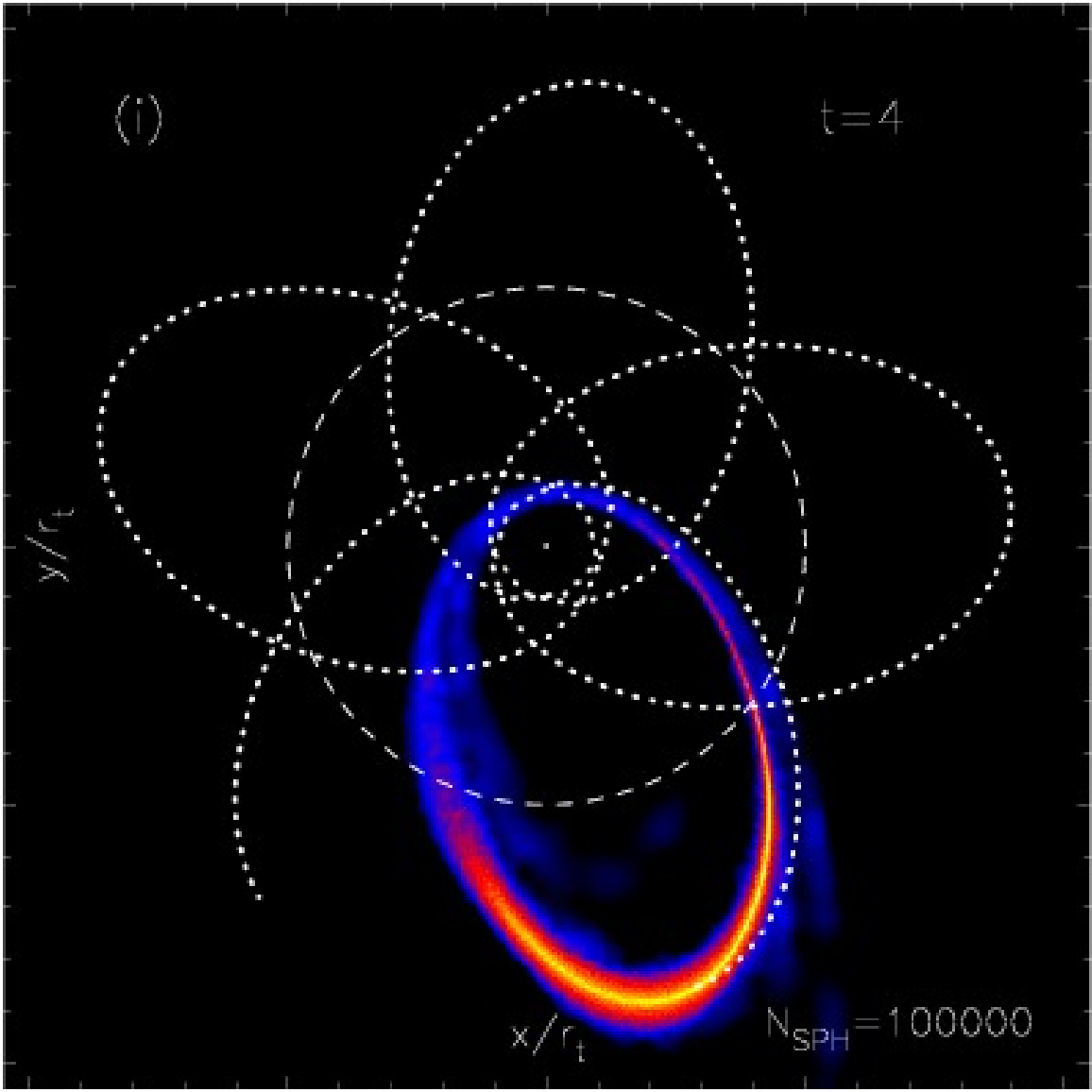}
}
\caption{Same format as Model~2a but for Model~2b.}
\label{fig:nodiskform}
\end{figure*}
%
Here we describe how an accretion disk forms rapidly around the
black hole in Model~2a.  The orbital angular momentum of a star
passing inside the tidal disruption radius should be conserved before
and after the tidal disruption, if there is no mechanism to
redistribute the angular momentum. The orbits of the stellar debris can
then be circularized by dissipation of orbital energy, primarily due to shocks from orbit crossing, which
conserve orbital angular momentum. Debris semi-major axes will thus
approach the circularization radius of the initial stellar orbit.

The specific energy and specific angular momentum of a test particle with the orbital 
parameters of Model~2a, moving under the pseudo-Newtonian potential, are given 
as $\epsilon_{\rm{tp}}=-0.497$ and $l_{\rm{tp}}=0.714$, respectively, by substituting 
$(r_{\rm{p}}, r_{\rm{a}})=(0.2r_{\rm{t}},1.8r_{\rm{t}})$ with $(e,\beta)=(0.8,5)$ into 
equations (\ref{eq:enam}) and (\ref{eq:wegg}).
Fig.~\ref{fig:eevol} shows the evolution of the specific energy, specific internal 
energy, and the specific angular momentum in Model~2a. In panel (a), the solid line 
and the dashed line show the SPH specific energy and that of a test particle, respectively. 
The small inside panel shows the evolution of the specific internal energy, which is
dissipated and radiated away after $t=2.4\Omega^{-1}$ 
due to the shocks from self-intersections of the debris orbits induced by relativistic precession.
Since the internal energy contributes negligibly to the total energy budget,
the specific energy approximately equals the specific binding energy.
On the other hand, panel (b) shows the evolution of the specific angular 
momentum normalized by $l_{\rm{tp}}$. Its marginal fluctuation 
is a numerical error with a magnitude of $0.15\%$, 
indicating that the specific angular momentum is almost conserved.
Assuming that the circularized disk has an axisymmetric 
Keplerian rotation, we can evaluate that the circularization radius $r_{\rm{circ}}$ 
is $\approx 2.5r_{\rm{p}}$.

Fig.~\ref{fig:diskform} shows sequential snapshots of the surface density of 
stellar debris (projected on the $x$-$y$ plane in five orders of magnitude, in a 
logarithmic scale) for Model~2a. The central small point, dashed circle, and dotted 
line show the black hole, tidal disruption radius, and orbits of a test particle 
moving under the pseudo-Newtonian potential, respectively. 
The run time is noted at the top-right corner in units of $\Omega_*^{-1}$, 
while the number of SPH particles are indicated at the bottom-right corner.
The star is tidally disrupted before it passes through the first pericenter at $t=0.5$.
Afterward, the stellar debris expands along the orbit of the test particle 
as shown in panel (a) and (b). It is stretched towards the second pericenter 
as shown in panel (c). At the second pericenter, the stretched debris weakly intersects 
with the remnant of the debris in panel (d) and then reaches the second apocenter 
while continuing to expand, in panel (e). When it reaches the third pericenter in panel (f), 
two stretched orbits clearly cross over. The orbital energy is significantly dissipated 
by the shock from orbit crossing between the two stretched debris streams. 
This is consistent with the decrease of specific energy in panel (a) of 
Fig.~\ref{fig:eevol}. The stellar debris is therefore rapidly circularized 
as shown in panels (g) and (h). Finally, in panel (i), a disk like structure is formed 
around the black hole sufficiently inside the tidal disruption radius.
Note that the number of SPH particles are slightly reduced from panel (e) to panel (i), 
since a very small fraction of the total number of SPH particles is accreted onto the black hole.
The trajectory of a SPH particle picked out of this simulation for $0\le{t}\la4$ 
is drawn in the solid line of Fig.~\ref{fig:grweggsim}. 
The size of the accretion disk is in rough agreement with $r_{\rm{circ}}\approx 2.5r_{\rm{p}}$.

Although the pseudo-Newtonian potential does not accurately model the
precession of near-circular orbits, we note that by the time the
pseudo-Newtonian potential becomes significantly inaccurate, the
debris streams have entered a regime of frequent orbit crossings,
guaranteeing further rapid circularization.  Furthermore, at a
qualitative level the pseudo-Newtonian potential probably
overestimates the circularization timescale, as it underestimates the
true rate of relativistic precession for eccentric orbits. However,
the pseudo-Newtonian potential still plays a crucial role in orbital
circularization processes, because the motion of a test particle under
the Newtonian potential take place in a closed path and therefore
causes no orbit crossing.  In order to test this, we have performed
the simulation for Model~2b, which has the same simulation parameters
as Model~2a except that the Newtonian potential is adopted.
Fig.~\ref{fig:nodiskform} shows sequential snapshots of the
surface density of stellar debris for Model~2b. It has the same format
as Fig.~\ref{fig:diskform}.  We note that there is no significant
evidence for orbital crossings during the timescale of the simulation,
since SPH particles orbit around the black hole on fixed eccentric
orbits. This is also confirmed by the fact that the percentage change of
specific orbital energy from t=0 to t=4 for Model~2b is less than
$0.4\%$.

\begin{figure*}
\resizebox{\hsize}{!}{
\includegraphics*[width=3mm]{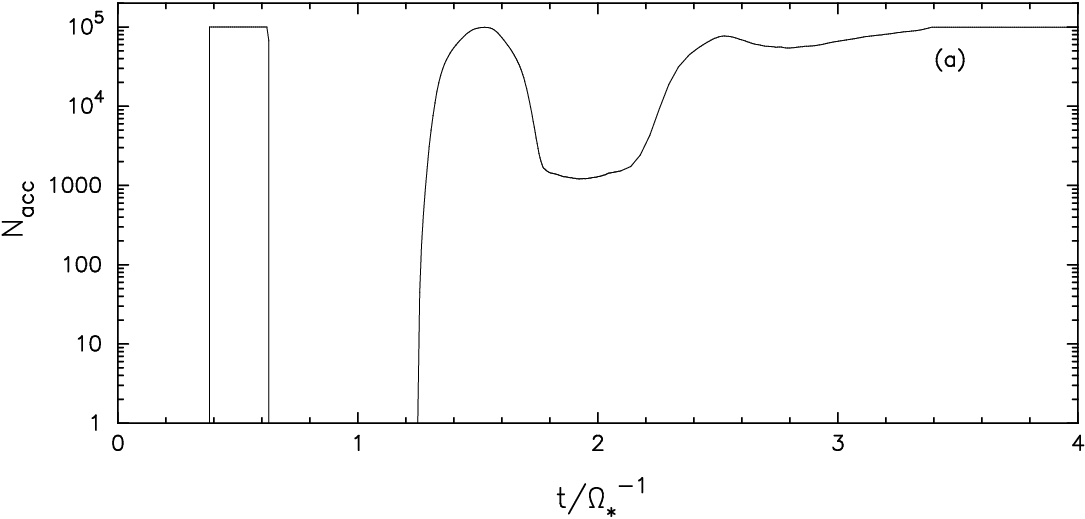}
}\\
\resizebox{\hsize}{!}{
\includegraphics*[width=3mm]{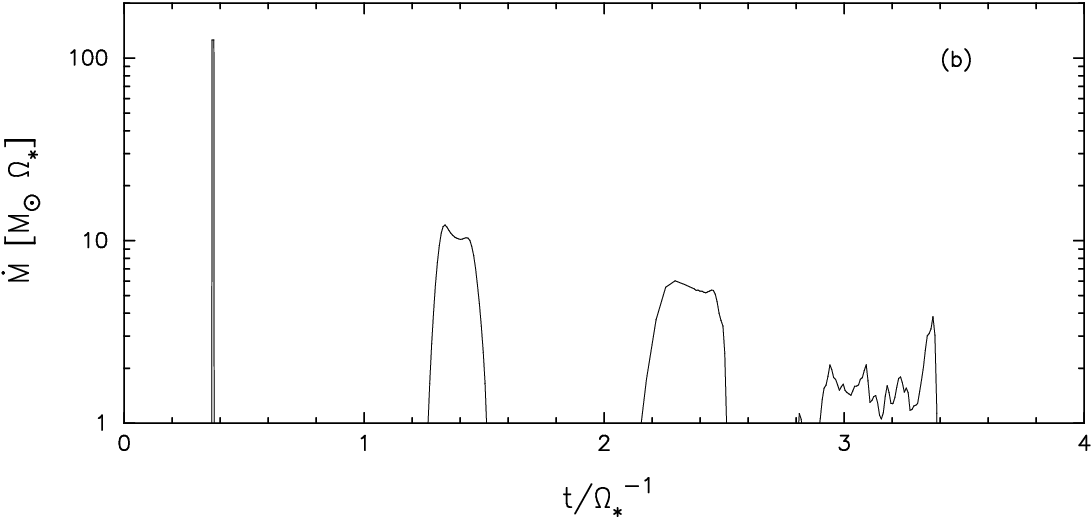}
}
\caption{
The number of SPH particles $N_{\rm{acc}}$ entering 
inside the tidal disruption radius over $0\le{t}\le4$ in panel~(a) 
and its corresponding mass capture rate, which is 
measured by the number of particles entering inside 
the tidal disruption radius per unit time.
In panel (b), the mass capture rate is normalized by 
$M_\odot/\Omega_{*}^{-1}$, which corresponds to 
$8.9\times10^5\dot{M}_{\rm{Edd}}$ for a $10^6M_\odot$ 
black hole with mass-to-energy conversion efficiency 
$\eta=0.1$.
}
\label{fig:mdot_tdr}
\end{figure*}

Fig.~\ref{fig:mdot_tdr} shows the time-dependence of the number of
SPH particles inside the tidal disruption radius $N_{\rm{acc}}$ and
its first derivative (the mass capture rate).  After the tidal
disruption of the star, its orbit passes through the first
apocenter, going completely outside of the tidal disruption radius.
The first peak of $N_{\rm{acc}}$ in panel~(a) comes when the debris
streams pass from the first apocenter to the second apocenter via the
second pericenter, and the stretched debris re-enters the tidal
disruption radius.  Part of it exits once more, but the fractional
remaining part is still inside the tidal disruption radius.  A
sequence of these events can be seen in panels (c)-(e) of
Fig.~\ref{fig:diskform}.  The second peak of $N_{\rm{acc}}$ forms in
panel~(b). The stretched debris returns again to the tidal disruption
radius, moving toward the third pericenter.  Afterwards, most of debris
circularizes and remains inside of the tidal disruption radius.

\begin{figure}
\resizebox{\hsize}{!}{
\includegraphics{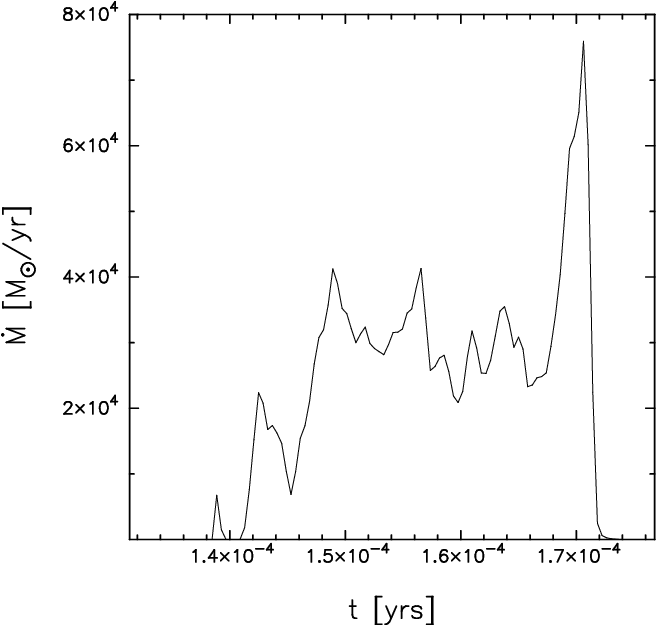}}
\caption{Mass accretion rate normalized by 1 solar mass per year, 
which corresponds to the final peak in panel (b) of Fig.~\ref{fig:mdot_tdr}.
}
\label{fig:mdot_disk}
\end{figure}

Panel~(b) of Fig.~\ref{fig:mdot_tdr} shows the rate of mass being
captured inside the tidal disruption radius. The first three peaks are
formed as stellar debris passes in and out the tidal disruption
radius, while the final peak shows the mass transfer rate to the
accretion disk around the black hole. The mass transfer rate is also
clipped from panel~(b) to Fig.~\ref{fig:mdot_disk}, as the mass
accretion rate normalized by 1 $M_\odot~{\rm yr^{-1}}$. It is clear
that the mass accretion rate has completely deviated from not only the
canonical $t^{-5/3}$ law but also the accretion rate variation of the 
other highly eccentric TDEs ( see Panel (b) of Figs. \ref{fig:model3a} and \ref{fig:model3b}).
The resultant accretion rate is more than five orders of magnitude
higher than the Eddington rate (see equation~\ref{eq:eddrate}).
This shows that the accretion flow is extremely supercritical in the case of moderately eccentric TDEs.

Since the viscous timescale measured at $r_{\rm{circ}}$ is estimated 
to be $t_{\rm{vis}}\sim3.5\times10^3(0.1/\alpha_{\rm{SS}})\Omega_*^{-1}$ 
where $\alpha_{\rm{SS}}$ is the Shakura-Sunyaev viscosity parameter, 
the accretion flow is clearly in a non-steady state.
Here,  we assume that the accretion disk is a geometrically thick
: $r_{\rm{circ}}/H\sim1$, where $H$ is the disk scale height.
Although the fate of the circularized debris is unclear because of much shorter simulation time
than the viscous timescale, the super-Eddington accretion flow will likely drive a
powerful outflow \citep{omnm05} as it becomes radiation-pressure dominated. 
This may increase the optical luminosity of the flare by orders of magnitude \citep{sq09}.  
It is even possible that a radiation-pressure supported envelope could be formed \citep{lu97}.

%
%

%
\section{Sources of Eccentric TDEs}
\label{sec:rates}
%

Although the canonical source for TDEs is two-body scattering at
parsec scales, a variety of alternate mechanisms exist.  In this
section we briefly review these, and highlight those which could
supply stars at eccentricities $e<e_{\rm crit}$ to their central SMBH.
\begin{itemize}

\item {\bf Resonant relaxation}: at small spatial scales close to a
SMBH, the approximately Keplerian potential causes stellar angular
momentum to relax in a correlated, coherent way (as opposed to the
uncorrelated changes from two-body relaxation).  It has been suggested
that the rapid changes in orbital eccentricity produced by resonant
relaxation (RR) may enhance the TDE rate at small scales \citep{ri98};
however, recent work has indicated that the complicated interplay
between RR and general relativistic precession gives rise to a
``Schwarzschild barrier,'' which quenches RR for stars with small
semi-major axis \citep{mamw11}.  RR is therefore unlikely to produce
TDEs with $e<e_{\rm crit}$.

\item{\bf Nuclear triaxiality}: conversely, the triaxial, non-Keplerian potential of a surrounding star cluster results in fewer constants of motion for stellar orbits; in particular, angular momentum is not conserved and some stars can wander arbitrarily close to the SMBH, eventually being disrupted \citep{mp04}.  However, this effect only arises at large radii where the potential from the star cluster can induce a significant triaxial correction to the SMBH potential, and is therefore unlikely to produce low eccentricity TDEs.

\item {\bf A nuclear stellar disk}:
there is a rotating stellar disk composed of young massive stars 
at the center of the Milky Way \citep{bmt10}.  
These stars orbit with moderate eccentricity around the supermassive black hole $\rm{SgrA^*}$.  
The nonresonant relaxation timescale drops as a star cluster becomes flattened and disk-like, due to greater encounter rates between stars; simple estimates of the two-body eccentricity relaxation timescale in the Milky Way's stellar disk ($\sim 5000 M_{\odot}$) find values $t_{\rm rlx}\sim4\times10^8$ yrs  \citep{kt11}.  Assuming a typical stellar mass of $10 M_{\odot}$ gives a disk star TDE rate of $\sim 10^{-6}~{\rm yr}^{-1}$, implying that if analogous disks are common in other galaxies, they could contribute nontrivially to the total TDE rate.  However, in order to produce TDEs with $e<e_{\rm crit}$, disk stars would need to be scattered in from $\sim~{\rm mpc}$ scales, implying that more compact disks than the Milky Way's are needed to produce significantly eccentric TDEs.

\item{\bf Binary SMBHs}: A binary SMBH system will, for a period of
$\sim 10^5~{\rm yr}$ to $\sim 10^6~{\rm yr}$, see a TDE rate
enhancement up to $10^{-1}~{\rm yr}^{-1}$ from a combination of
chaotic orbital evolution and Kozai cycles \citep{ips05,
cms09}. Recent work suggests that chaotic orbits in particular are the
dominant contribution to the rate enhancement \citep{cms11, wb11}.
Most of the stellar flux originates from within one order of magnitude
of the semimajor axis of the binary; since the inspiral of binary
SMBHs stalls when the lower-mass secondary reaches a radius inside of
which is contained its own mass in stars, a low binary mass ratio $q$
will produce a flux of less eccentric TDEs.  In particular, Fig.~17 of
\cite{cms11} indicates that for $q=1/81$ and a primary black hole mass
of $10^7 M_{\odot}$, some stellar flux into the
primary's loss cone originates from spatial scales $\le 10^{15}\,{\rm
cm}$, which implies $e<e_{\rm crit}$ even for $\beta=1$ events.
However, the chaotic orbits which dominate the rate enhancement
produce TDEs sampling a wide range of $\beta$ \citep{mp04}, so the
situation is even more favorable. Binary SMBHs on a eccentric orbit
appear to produce even more TDEs from small apocenters.  The
``hardening radius'' (the orbital radius within which the stellar cusp
of the primary, larger SMBH contains the mass in stars of the
secondary, smaller SMBH) effectively sets the scale at which stellar
disruption rates are enhanced; to produce eccentric TDEs requires
relatively low mass ratios ($<1/50$) which could in some cases be
characterized as SMBH-IMBH inspirals.  Because binary SMBHs are
expected to provide up to $10\%$ of the total TDE rate \citep{wb11},
it is important to note that a subset of these events may deviate from
canonical ``parabolic'' light curves.

\item{\bf SMBH recoil}: the gravitational wave recoil accompanying a
SMBH merger will scramble the orbits of surrounding stars and
partially refill the kicked merger remnant's loss cone, briefly
increasing the TDE rate.  If one considers the bulk of the stellar
population surrounding the merging binary black holes, the excavation
of a ``binary loss cone'' results in too few stars at small separation
for the resulting burst of TDEs to involve any events with $e<e_{\rm
crit}$ \citep{sl11}.  However, at low binary mass ratios, it is
possible for mean motion resonances to migrate stars to small
semimajor axes during the binary inspiral \citep{schnittman10, sm10,
sm11}; any TDEs produced from this stellar subpopulation would have
$e<e_{\rm crit}$, and would be described by this paper.

\item{\bf Binary separation:}
Recent theoretical studies on rates of tidal separation of binary
stars by SMBHs suggest that a significant fraction of tidal disruption flares 
may occur from stars approaching the black hole from subparsec scales \citep{asp11, bcb12}. When a binary star passes 
sufficiently close to the black hole to be tidally separated (without immediate 
disruption of either component), one star becomes tightly bound to the black hole 
while the other is flung away at a high speed.  The subsequent orbital evolution of the bound star will represent a competition between gravitational radiation and stellar relaxation processes. If gravitational radiation dominates, the orbit will circularize and spiral in, likely leading to a phase of steady mass transfer that is unlike the eccentric and violent disruptions which we have simulated in this paper.  On the other hand, if stellar relaxational processes dominate, the star can eventually diffuse into the loss cone with a nonzero eccentricity that is still significantly smaller than that expected for TDEs generated by two-body relaxation at parsec scales.  For a bound star with $r_{\rm p}>r_{\rm t}$ and semimajor axis $a$, living in a stellar cusp with density profile $\rho(r) \propto r^{-b}$, the eccentricity dividing gravitational-wave-dominated evolution from relaxation-dominated evolution is \citep{asp11}
\begin{eqnarray}
e_{\rm gw} \approx &1-0.016\times(8\times 10^{-4})^{(2b-3)/5}(3-b)^{2/5}\label{aspGW} \\
&\times \left(\frac{M_{\rm BH}}{10^7M_{\odot}}\right)^{(8-b)/5} \left( \frac{a}{1000~{\rm AU}} \right)^{(2b-11)/5} \nonumber
\end{eqnarray}
In practice, requiring $e_{\rm gw}>e_{\rm crit}$ for $a<r_{\rm
t}/(1-e_{\rm crit})$ is only possible for steep stellar cusps
$b\approx 2$ and very low mass SMBHs ($M_{\rm BH} \le
10^{5.5}M_{\odot}$).  A further complication is the presence of the
Schwarzschild barrier, which may prevent stars with long gravitational
wave inspiral timescales from entering the loss cone.  Stars bound to
a SMBH from binary separation events can become TDEs with lower
eccentricity than in the canonical scenario, but equation~(\ref{aspGW}) implies their eccentricity
will still tend to be greater than $e_{\rm crit}$.  On the other hand, equation~(\ref{aspGW}) was derived assuming the Fokker-Planck diffusion limit for two-body relaxation, and anomalous diffusion from strong scattering events \citep{bka12} may allow violent tidal disruption of separated binaries even from eccentricities $e<e_{\rm gw}$.

\end{itemize}

Of the possibilities we have considered, two stand out as particularly promising:
TDEs generated as binary SMBHs harden and stall, and TDEs generated
during the coalescence of binary SMBHs with a low mass ratio by stars
that were brought inward via mean motion resonances.  The first
possibility may account for a significant fraction of the total TDE
rate; the second possibility will be much less intrinsically common,
but could serve as an electromagnetic counterpart to an
eLISA/NGO\footnote{http://www.elisa-ngo.org/}-band gravitational wave
signal.  It is also possible that strong scatterings could lead to eccentric TDEs from the bound stars produced by tidal separation of stellar binaries.  Other possible enhancements to theoretical TDE rates, such as
perturbations from infalling giant molecular clouds, likely occur at
too large of a spatial scale to create eccentric TDEs.

\section{Summary \& Discussion}
\label{sec:sumdis}
%
We have carried out numerical simulations of mass fallback and accretion processes
around a SMBH.  Specifically, we have examined the tidal disruption of a star on a bound 
orbit, considering relativistic effects with a pseudo-Newtonian potential. 
Using a polytropic gas sphere as our initial conditions, we have considered 
both parabolic orbits ($e=1.0$) and eccentric ones ($e = 0.8$ and $0.98$), varying the 
penetration factor $\beta$ as well.

We have found that a non-steady, non-axisymmetric accretion disk is
formed around the black hole in the case of e=0.8 and $\beta=5$. The
formation of an accretion disk occurs as follows: a segment of the
stellar debris returning to pericenter, and a different one exiting
pericenter, intersect each other due to relativistic precession. The
orbital energy is then dissipated by shocks due to orbit crossing.
Since the orbital angular momentum of the stellar debris is conserved
before and after the tidal disruption, the debris orbits are rapidly
circularized during a few orbit crossings.  This shows that the
initial size of the accretion disk is only determined by the orbital
angular momentum of the initial stellar orbit. In our simulations, the
circularization radius is estimated as
$r_{\rm{circ}}\approx2.5r_{\rm{p}}$ where $r_{\rm p}$ is the
pericenter distance. Furthermore, the expected accretion rate is
extremely super-Eddington.

The striking difference between moderate-eccentricity simulations with and
without relativistic precession highlights the importance of general
relativistic effects for debris circularization.  Specifically, very
little energy dissipation was seen in Model~2b (Newtonian) of Table
1, while its pseudo-Newtonian equivalent, Model~2a, saw rapid
accretion disk formation.  For reasons of computational cost, we did
not follow the longer timescales required for debris circularization
in other models, but in future work we hope to investigate if these
results generalize to higher eccentricity and parabolic TDEs.
Sufficiently rapid SMBH spin could delay disk formation, as
Lense-Thirring torques will precess the orbital planes of individual
debris streams and limit or prevent orbit-crossings. If a disk is able
to eventually form, its luminosity will be periodically modulated by
Lense-Thirring precession (e.g., \citealt{sl12a}).  Even considering
this complication, the energy dissipation process is crucial for
determining the formation and structure of an accretion disk. In these
exploratory simulations, we have adopted the polytropic equation of
state instead of solving an energy equation. In a subsequent paper, we
will study the detailed disk formation and structure by solving more
realistic energy equations with and without a radiative cooling term.

Eccentric TDEs are a subpopulation of all TDEs, but a potentially
interesting one.  The two most promising means of producing them are
in the dynamical friction stage of binary SMBHs, and
immediately following gravitational wave-driven black hole coalescence. 
Both of these are especially interesting subsets of TDEs: the former occurs
during a time of greatly enhanced TDE rates, when it may be possible
for a single galaxy to produce multiple TDEs in $\sim 10$ yr, while
the latter would serve as a delayed electromagnetic counterpart to a
low-frequency gravitational wave signal.  It is also possible that
tidal separation of binary stars could produce TDEs of interestingly 
low-to-moderate eccentricity, if a large population of $M_{\rm BH} \le
10^{5.5}M_{\odot}$ SMBHs reside in steep stellar cusps.  Eccentric
TDEs are also of interest because of the relatively short delay time
between disruption and disk formation, allowing numerical simulations
to bypass costly apocenter passages and directly approach open
problems in debris circularization.  The pseudo-Newtonian potential
proposed by \citet{wc12} is, however, not applicable for the very low
eccentricities of fully circularized gas because it was derived with
the assumption of small binding energy.  Therefore, we hope to employ
post-Newtonian corrections in a subsequent paper.

For eccentric stellar orbits, there is a critical value of the orbital
eccentricity below which all stellar debris is bounded to the black
hole. It can be seen from our simulations that the critical
eccentricity is slightly lower than expected from our analytical
predictions. This might be because of the effects of stellar structure
on the tidal disruption. 
There are three important implications for lightcurves of eccentric TDEs:
\begin{itemize}
\item The mass fallback rate will not asymptote to $\dot{M} \propto t^{-5/3}$ 
at late times but be finite with cut-off time $t=t_{\rm max}$.
\item The lack of unbound debris will eliminate observational signatures 
associated with emission lines (e.g., \citealt{sq09}).
\item A larger amount of mass will return to pericenter in a much shorter time 
than in the standard parabolic picture, considerably increasing the ratio of 
$\dot{M}/M_{\rm Edd}$ for the subsequent flare.
\end{itemize}
Even for eccentricities $e>e_{\rm crit}$, the center of the
differential mass distribution will shift in a negative direction,
providing weaker versions of the above effects. These signatures
should be searched for when large samples of TDE candidates from next
generation optical transient surveys such as the Large Synoptic Survey
Telescope\footnote{http://www.lsst.org/lsst/} and next generation all-sky X-ray surveys
such as extended Roentgen Survey with an Imaging Telescope Array\footnote{http://www.mpe.mpg.de/eROSITA} 
become available.

%
%
%

%
\section*{Acknowledgments}
%
KH is grateful to Atsuo.~T Okazaki, Takahiro Tanaka, and Shin Mineshige 
for helpful discussions and their continuous encouragement.
The numerical simulations reported here were 
performed using the computer facilities of Yukawa Institute of Theoretical Physics, 
Kyoto University and Harvard-Smithsonian Center for Astrophysics, Harvard University.
 This work was supported in part by the Grants-in-Aid of the Ministry of 
Education, Science, Culture, and Sport and Technology 
[21540304, 22340045, 22540243, 23540271 KH], 
Education and Research Promotion Foundation in Kyoto University [KH], and 
NSF grant AST-0907890 and NASA grants NNX08AL43G and NNA09DB30A 
[NS, AL].
%

\end{document}